\documentclass[preprint,tightenlines]{revtex4-1}  
\usepackage[]{graphicx}
\usepackage{color}
\usepackage{bm}
\usepackage{amsmath}
\usepackage{amssymb}
\usepackage[makeroom]{cancel}
\usepackage{changes}
\usepackage{appendix}
\usepackage{xfrac}
\definecolor{purple}{cmyk}{0.1,0.9,0,0.1}
\definecolor{dgreen}{cmyk}{0.8,0,0.8,0.2}
\definechangesauthor[name={Vitaly}, color=dgreen]{vr}
\definechangesauthor[name={Sophya}, color=purple]{sg}
\definechangesauthor[name={Julian}, color=blue]{js}
\newcommand{\be}{\begin{equation}}
\newcommand{\ee}{\end{equation}} 
\newcommand{\bea}{\begin{eqnarray}}
\newcommand{\eea}{\end{eqnarray}} 
\newcommand{\bra}{\langle}
\newcommand{\ket}{\rangle} 
\newcommand{\grad}{\nabla} 
\newcommand{\pd}{\partial} 
\newcommand{\ba}{\begin{array}}
\newcommand{\ea}{\end{array}}

\begin{document}

\title{Analysis of divergent dynamics  of  exactly factorized  electron-nuclear  wavefunctions}

\author{Julian Stetzler}
\author{Sophya Garashchuk}
\author{Vitaly Rassolov}\email{rassolov@mailbox.sc.edu}
\affiliation{Department of Chemistry \& Biochemistry, University of South Carolina, Columbia, South Carolina 29208}

\begin{abstract}
The Exact Factorization (XF)  of molecular wavefunctions can be viewed as an 'electronic wavepacket'  framework for quantum dynamics.   It is an appealing alternative to the  conventional non-adiabatic dynamics, unfolding  in the space of   coupled  electronic eigenstates.   However,  implementation of the non-linear  XF equations  for general systems  presents a formidable challenge: the XF counterparts to the non-adiabatic coupling  involve division by the nuclear probability density, which  leads to severe numerical instabilities  in the low-density regions of space.  In case of the non-adiabatic dynamics  the effect of coupling is relatively smooth, but this theoretical framework becomes impractical  when numerous electronic states are involved.  In this paper  the origin of the  XF-specific challenge is analyzed analytically. We   demonstrate  that the problem arises  when the factorized wavefunction diverges,   even without the explicit coupling of the Born-Huang electronic states used to describe the molecular wavefunction.   Using a 'minimal' model of the photodissociation, we derive expressions for the XF dynamics and  locate the source  of the XF instability. We  analyze  the  dependence  of this instability on the nuclear wavefunction bifurcation  in the stationary  and moving frames of reference, the latter associated with the  quantum trajectory ensemble  describing the nuclear XF wavepacket in a compact form. We  show that the near-singular behavior  persists in the moving frame and in the atomic basis representation of the electronic wavefunction.   This model and  insight into the   root  of   the  XF implementation challenge will  help to  address the issue, leading to  further development of  the  XF methods.

\end{abstract}

\maketitle

\section{Introduction}
A growing number of experiments highlight the nuclear quantum effects on    the electron-nuclear dynamics of  large molecular systems, including chemical processes in metals and the condensed phase.   Some examples  include the coupling of  electronic excitations  to the phonon modes, observed in the vibrational relaxation of CO on Cu or NaCl \cite{TullyMetalsReview,srt2009}, and the formation of polarons in polarizable materials \cite{frs21,fww17}. For theorists the challenge of describing such processes lies in the failure of the Born-Oppenheimer (BO) approximation,  and the inefficiency of the Born-Huang (BH) expansion of molecular wavefunctions requiring  computation of  many electronic eigenstates \cite{TullyMetalsReview}. Recent experiments in polaritonic chemistry further  motivate  the development of theoretical   methods  for  strongly coupled quantum photon-electron-nuclear systems \cite{PolaritonReview}.     One such development   is  the exact factorization (XF) method, introduced to the field of theoretical chemistry in Ref. \cite{amg10}. The XF   is  an approach   to   quantum molecular dynamics based on the product form of time-dependent multi-component wavefunctions, which for clarity of discussion here  will be taken  as the electron-nuclear wavefunction,  
    $$  \Psi(x,y,t) = \underbrace{\psi(y,t)}_{nuclear}\underbrace{ \Phi(x,y,t)}_{electronic}, $$
   subject to the partial normalization condition   of $\Phi$ in the electronic space its coordinate being $x$. This condition  is defined as  $\bra\Phi|\Phi\ket_x=1$ for any configuration of the nuclei (coordinate   $y$) at all times, $t$.  Note that, formally,  to be 'exact' the electronic component, $\Phi$, is a function of  both the electronic and nuclear coordinates. The  XF representation is expected to be advantageous for systems  involving  many electronic states with  $\Phi(x,y,t)$    describing    an electronic wavepacket, and as a starting point for well-defined approximations to the  dynamics of both wavefunction components, necessary for applications to large molecular systems. 

Recent conceptual developments include an XF-based theory of electronic friction \cite{RoccoIreneFriction}, the factorized electron-nuclear dynamics (FENDy) with a complex potential \cite{Fendy}, and extension to polaritonic chemistry \cite{NeepaPolariton,TokatlyPolariton,XFPhotonNuclear}. Furthermore, the  XF ideas were  used to develop the approximate Coupled Trajectory-Mixed Quantum Classical (CT-MQC) method \cite{CTMQC}. XF-derived decoherence corrections have also been developed for surface hopping and are included in the Libra software package for use with both model and atomistic systems \cite{XFLibra}. 

Thus far,  applications of the  'exact'  XF  were limited to  small model systems, such as Tully models \cite{XFTullyModels, Fendy}, 1-dimensional H$_2^+$ in a laser field with soft coulomb interactions \cite{ElectronLocalisation,XFH2LaserField}, and the Shin-Metiu \cite{sm95,ShinMetiuXF} model where accurate solutions on the full-space can be used.  In these studies, XF provided conceptual understanding of molecular Berry phases \cite{XFBerryPhase},  and the forces driving nuclear dynamics in non-adiabatic systems \cite{BasileAgostiniSteps,XFSteps}. Perhaps,  the largest chemical   application to date  is the study of the ring-opening of oxirane, where the approximate CT-MQC method was applied to a realistic molecular system using on-the-fly Density Functional Theory computation of the electronic structure \cite{CTMQCApplied,XFOxirane,IbeleNamdReview}. 

Exact numerical implementation of XF presents a formidable   challenge, 
  analyzed  in two notable studies:  a rigorous mathematical investigation by Lorin \cite{LORINXF} and a direct numerical algorithm developed by Gossel and co-workers \cite{NeepaNumericalXF}. These studies mainly attribute the numerical difficulties to the non-classical, also referred to as "quantum", momentum:
   $$ r(y,t):=\frac{\grad_y|\psi(y,t)|}{|\psi(y,t)|}.$$ 
 Based on the ongoing research in our group, we argue that the spatial separation of the nuclear wavefunction plays a key role in the numerical stability of the dynamics (Unpublished Observation). As the separation of the diverging wavepackets leads to the near discontinuous steps in the TDPES observed in Ref. \cite{XFSteps} and will occur regardless of the  non-adiabatic coupling. The nuclear separation has been identified as the cause of these steps \cite{BasileAgostiniSteps}. 
  In this work we  investigate this regime of dynamics analytically using a simple model which can be used for future method development.
    The remainder of this paper is organized as follows: Section \ref{sec:theory}  presents the model and theoretical background, Section \ref{sec:TheoryAnalysis} presents the analytic results and analysis. Section \ref{sec:summary} concludes.

\section{The Formalism and Model}\label{sec:theory}

Henceforth we will  use the  atomic units ($\hbar=1$) and  work  with  rescaled  system  parameters  defined in Table \ref{tab:Definitions}. We will refer to the light and heavy particle degrees of freedom (DOFs)  as 'electronic' and 'nuclear',  and assume the electronic mass of $m=1$ a.u.;   $M$ is the nuclear mass, $ M\gg m$.  For simplicity of the analysis,  we will consider a two-dimensional system  using $x$  and $y$ to  denote the  electronic   and nuclear  DOFs, respectively.  
   The spatial derivatives will be labeled as gradients, i.e.  $\grad_y$=$\frac{\pd}{\pd y}$,  
    while 
   $\frac{\pd}{\pd t}$  denotes the  partial time-derivative in the stationary, or Eulerian, frame, and  
   $\frac{d}{dt}$  denotes the full time-derivative in the moving, or Lagrangian, frame of reference. The time-dependent parameters are indicated  by  the  subscript $t$.  Integration over $x$ only is denoted $\bra...\ket_x$. Arguments of functions are dropped when unambiguous.

\subsection{Exact factorization of the molecular  wavefunction } 

First, let us summarize  the  exact factorization (XF)  formalism, specifically the  factorized electron-nuclear dynamics (FENDy)  with complex potential, presented in Ref. \cite{Fendy}.   A notable difference between  FENDy and the XF framework introduced by  Abedi, Maitra and Gross \cite{amg10}, is that  the former  is  based on the complex scalar potential rather than on the vector potential. 

Within the  XF  formalism  the electron-nuclear wavefunction $\Psi(x,y,t)$  is represented as a product of the purely {\it nuclear} component, $\psi(y,t)$,   and the {\it electronic} component,  $\Phi(x,y,t)$,    
\be \Psi(x,y,t) = \psi(y,t) \Phi(x,y,t),\label{eq:XFWF} \ee 
 where $\Phi(x,y,t)$ satisfies the  partial normalization condition,  
\be \bra \Phi|\Phi \ket_x = 1 \mbox{~for~all~} y \mbox{~and~} t.  \label{eq:normalization}\ee
For such representation to be exact, the electronic function formally depends on both $x$ and $y$. The goal is to have such factorization of the molecular wavefunction, $\Psi$, that  'most' of the dynamics is captured by  $\psi(y,t)$, which will simplify the time-dependence  of $\Phi(x,y,t)$ and  make it amenable to approximations.  Adding and  subtracting a complex  time-dependent potential energy surface (TDPES) $V_d(y,t)$,  the molecular  time-dependent Schr{\"o}dinger equation (TDSE) is separated into the  nuclear  and electronic equations, respectively: 
\be \hat{K}_y\psi+V_d(y,t)\psi=\imath \frac{\pd\psi}{\pd t} \label{eq:nucTDSE},\ee
\be \hat{H}^{BO}\Phi+(\hat{D}_2+\hat{D}_1)\Phi +\hat{V}_{ext}\Phi -V_d(y,t)\Phi=\imath \frac{\pd\Phi}{\pd t} \label{eq:eTDSE}.\ee 
The nuclear TDSE, Eq.  (\ref{eq:nucTDSE}),   does not include  any  explicit coupling terms, only the nuclear kinetic energy $\hat{K}_y$ and the so-far unspecified complex TDPES, $V_d$, 
\be V_d(y,t):=V_r(y,t)+ \imath V_i(y,t). \label{eq:Vd} \ee
In the electronic TDSE (\ref{eq:eTDSE})  $\hat{H}^{BO}$ is the Born-Oppenheimer (BO) Hamiltonian consisting of  the  electronic kinetic energy and all Coulomb interactions.  If present,  the external field, e.g. the laser pulse, is included via $\hat{V}_{ext}$.  The TDSE (\ref{eq:eTDSE})   includes the coupling terms,   $\hat{D}_1$ and $\hat{D}_2$,  containing the derivatives of $\Phi$ with respect to the nuclear coordinate,
\be \hat{D}_1:=-\frac{\grad_y\psi}{\psi}\frac{\grad_y}{M} \label{eq:D1},\ee~
\be \hat{D}_2:=-\frac{\grad^2_y}{2M}. \label{eq:D2}\ee  The derivative coupling operator,   $\hat{D}_1$,  involves  
 the ratio $\grad_y \psi/\psi$  presenting   significant  challenges to the numerical implementation of  the  XF equations.   
 
Within  the  TDPES of Eq. (\ref{eq:Vd})  the purpose of   $V_r$ is to minimize the average residual nuclear momentum in $\Phi$,  while the role of $V_i$ is to maintain the partial normalization conditions (Eq. (\ref{eq:normalization})). As discussed in Ref. \cite{Fendy}, for a system with  a {\it single} nuclear dimension,  it is possible to factorize the full wavefunction exactly, such that both the imaginary potential,  $V_i$,  and the average residual nuclear momentum,  $\bra \Phi| \grad_y \Phi\ket$,  are equal to zero;  the real potential,  $V_r$, driving the evolution of the nuclear wavefunction,  $\psi(y,t)$,  is defined up to a time-dependent constant.    
However, in  our 'minimal'   model, described in Section \ref{sec:model},  
 we will not  use  $V_d$ or the electronic wavefunction explicitly.   We will simply  take  the electronic wavefunction  as an expansion in a basis of  $N_s$ eigenstates of the BO Hamiltonian,  $\{\phi_i(x,y)\}$,  
\be \bra \phi_i |\hat{H}^{BO}| \phi_j\ket_x = V_{ij}(y), \label{eq:BOstates}\ee
\be \Phi(x,y,t)=\sum_{i=1}^{N_s} C_i(y,t)\phi_i(x,y).  \label{eq:eansatz} \ee 
Thus, the  time-evolution of the electronic wavefunction $\Phi$  is described by the complex coefficients,  $\{C_i(y,t)\}$, which  will be analyzed  in terms of their  amplitudes and phases as functions of the nuclear position. The dynamics of  these coefficients will follow from the properties of the nuclear wavefunction represented in terms of the quantum trajectories reviewed in Section \ref{sec:QTs}.  

\subsection{The quantum trajectory formalism}\label{sec:QTs}
 Within the XF formalism, we want the key features of the nuclear dynamics  to be captured by $\psi(y,t)$.   For an efficient  in high-dimensions  representation of $\psi(y,t)$, we are considering $\psi(y,t)$    represented as an ensemble of interacting 
 Quantum Trajectories (QT), or Bohmian trajectories,  \cite{bohm52}.  The QT representation of wavefunctions 
  as a numerical method   received the attention of  theoretical chemists over the last 25 years (see e.g. Refs   \cite{wyatt_book,Bohmian,meier2023,Dupuy22a,Lombardini24}).  The  concept of QTs  is also  employed for interpretation of the QM phenomena \cite{sanz07,angel2024}, and  in a variety of approximate or semiclassical trajectory-based methods. The review  of the QT-inspired dynamics  is beyond  the scope of this work, but to name a few,  these methods  include dynamics with  linearized quantum force \cite{garashchuk04}, quantized Hamiltonian dynamics \cite{prezhdo06,akimov2018},   the QT-surface hopping methods (with and without the XF)  \cite{QTSH,AkimovQTSH,XFTullyModels}, and the QT-guided adaptable gaussian bases \cite{QTAG}.  There also Bohmian extensions into the phase space \cite{bittner02b}, multi-polar Bohmian dyanmics \cite{poirier08b} and stochastic Bohmian mechanics \cite{salvador2019}. Formal counterparts to the QT dynamics in non-adiabatic systems  have been also  proposed (e.g. Refs \cite{wyatt01a,Dupuy22b,garashchuk06b}). 
  
 In its original form,   an ensemble of the   QTs,  governed by  the Newtonian-like  equations of motion,   is used as  a formally exact representation of the wavefunction. The  basic formalism is outlined below, and  the detailed derivation  can be found, for example, in Ref.   \cite{wyatt_book}. 
 The QT formalism hinges on the polar representation of a wavefunction in terms of real phase, $S$, and amplitude, $|\psi|$,  
 \be \psi(y,t):=|\psi(y,t)|e^{\imath S(y,t)}, \label{eq:QTWF} \ee  
 substituted into the TDSE. 
 The gradient of the wavefunction phase,  $P(y,t)$,   
 \be P(y,t):=\grad_yS(y,t) \label{eq:QTMOM}, \ee  
 is  identified as  the momentum $p_t$ of a QT at the position  $y_t$,   \be p_t = P(y,t)|_{y=y_t}.\ee    Assuming here a one-dimensional cartesian  Hamiltonian, $\hat{H}=-\frac{\grad^2_y}{2M}+V(y)$, 
 the QT position changes  according to the classical  equation of motion, 
 \be \frac{dy_t}{dt}= \frac{p_t}{M}.\ee
 The  particle velocity, $p_t/M$,  defines the Lagrangian frame of reference, 
 \be \frac{d}{dt}:=\frac{\pd}{\pd t}+\frac{p_t}{M}\grad_y.   \label{eq:LagFrame} \ee
 The remaining QT equations following from the TDSE in the Lagrangian frame are:  
\bea \frac{dp_t}{dt} &=& -\grad_y(V+U)|_{y=y_t}\\
\frac{dS_t}{dt}&=&\frac{p_t^2}{2M}-(V+U)|_{y=y_t}, \label{eq:dSdt}\eea
where $U$ is the quantum potential incorporating quantum-mechanical behavior into otherwise classical equation of the trajectory motion,    
\be U(y,t):=-\frac{\hbar^2}{2M}\frac{\grad^2_y|\psi|}{|\psi|}.\ee
The complementary to Eq. (\ref{eq:dSdt})  continuity equation for the probability density, $\rho(y,t)=|\psi(y,t)|^2$,  leads to    the following useful property of the QT dynamics:  the weights, $w(y_t)$,  of QT trajectories,  defined as the probability density within the volume element $dy_t$ associated with the trajectory $(y_t,p_t)$,   
\be w(y_t)=|\psi(y_t)|^2dy_t. \label{eq:weight} \ee
are  constant in time \cite{garashchuk03}.  In one nuclear dimension this property allows one to generate  the QT position  from  the probability density of the full wavefunction  by simply  finding  $y_t$ for a fixed value of the  cumulative density, $\rho_c$, at every moment of time, 
\be \rho_c(y_t) = \int_{-\infty}^{y_t}\!\! |\psi(y,t)|^2 dy =  \int_{-\infty}^{y_t}\int_{-\infty}^\infty\! \!|\Psi(x,y,t)|^2 dx dy   \label{eq:rhoc}.\ee 
In Eq. (\ref{eq:rhoc})  the second equality applies for the XF wavefunction given the conditional normalization of Eq. (\ref{eq:normalization}).  Eq. (\ref{eq:rhoc}) is used to generate QTs numerically. {Additional  discussion and examples   of the QT dynamics    on a  single BO surface and   within the  XF  formalism  applied to a two-state system can be found  in Appendix \ref{sec:QTappendix}. 


\subsection{The model}\label{sec:model}
Our basic  model of   photodissociation,    
sketched in Fig. \ref{fig:Model}(a), consists of  the nucleus of mass $M$ described by the coordinate $y$, evolving on two electronic states, $N_s=2$.  We assume that a certain  fraction  of the nuclear wavepacket  was instantaneously transferred from the ground electronic state, $|\phi_1\ket$, to the excited electronic state,  $|\phi_2\ket$.    The electronic states are assumed to be orthonormal. 
\begin{figure}
    \centering
    \includegraphics[width=0.9\textwidth]{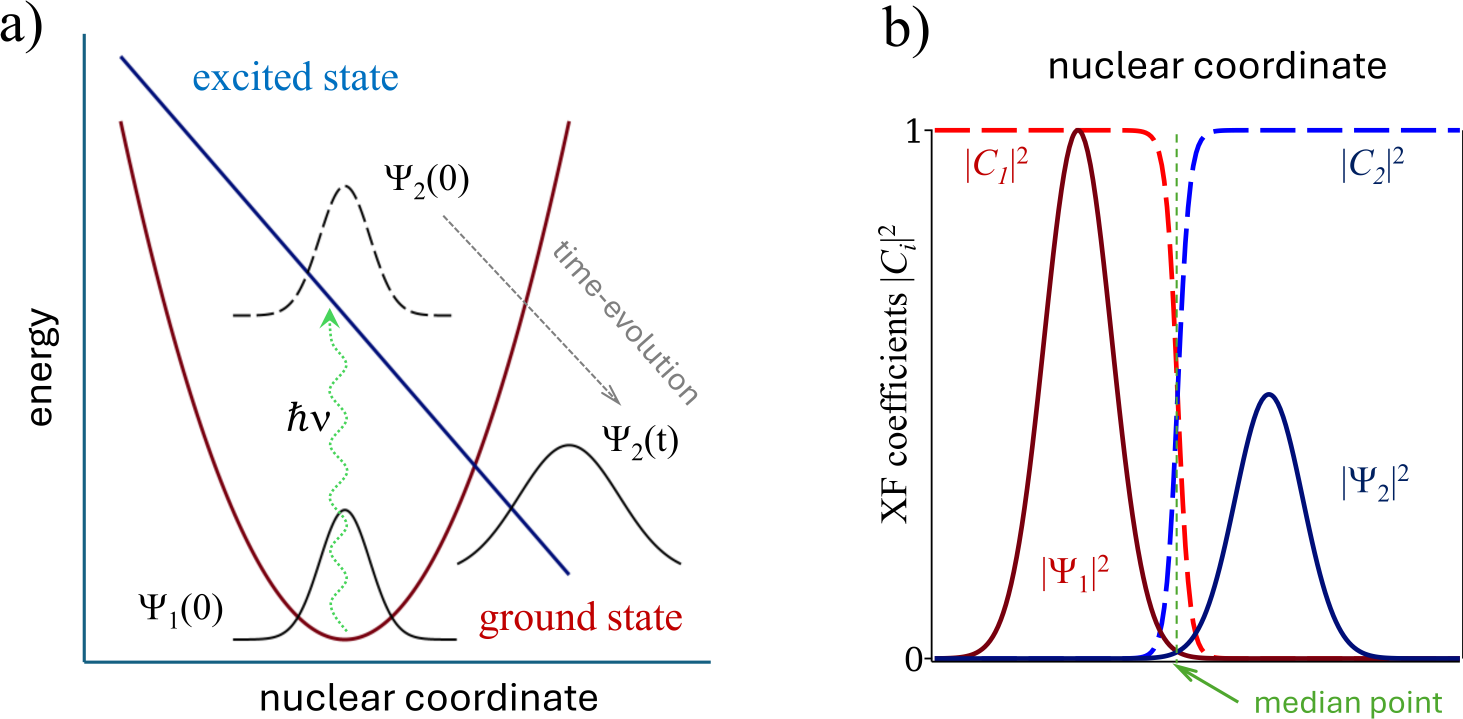}
   \caption{A minimal model of photodissociation dynamics based on two electronic states.  (a) A fraction of the  nuclear wavepacket, $\psi_1(0)$,  in the ground electronic state (red parabola) is instantaneously transferred  into the  excited electronic state (blue linear function)  of dissociative character. With time this wavepacket, $\psi_2$,  diverges from $\psi_1$ in the nuclear coordinate. (b)  The probability densities of the Born-Huang nuclear wavepackets $\psi_1$ (red solid line)  and $\psi_2$  (blue solid line) add up to the XF nuclear wavefunction, $|\psi|^2=\sum_i|\psi_i|^2$,  defining  the expansion  amplitudes, $|C_1|$  (red dash)  and $|C_2|$ (blue dash) of the XF complement to $\psi$ in the electron-nuclear space.
   For diverging wavepackets,  $|C_i|$ behave as step-functions near  the median point of equal $|\psi_i|^2$ values.}  
    \label{fig:Model}
\end{figure}  

In the ground electronic state, the  nuclear subsystem is described by  a  harmonic oscillator with the frequency $\omega$.  For   the  excited electronic state, the potential  is a linear function of $y$. The electronic states are   
uncoupled, i. e.  in  Eq. (\ref{eq:BOstates})   $V_{12}=0$.    Shifting  the potentials  by  $-\omega/2$  to reduce the time-dependence of the nuclear wavefunction in the ground state,    the Born-Oppenheimer Hamiltonian matrix in atomic units  becomes: 
\be 
{\mathbf H}^{BO}:=\begin{bmatrix}
    \frac{M\omega^2}{2}y^2-\frac{\omega}{2} & 0 \\
    0 & -k y -\frac{\omega}{2}
\end{bmatrix}.  
\ee    This model allows us to study the behavior of the XF electronic expansion coefficients $C_i(y,t)$ during  the nuclear dynamics without explicit functional form of the electronic functions.


\begin{table}
    \centering
    \begin{tabular}{c|c|c|c}
    \hline
    \multicolumn{4}{c}{XF nuclear wavefunction $\psi(y,t)$} \\\hline
    A & phase: $S:=\arg(\psi)$ & momentum: $P:=\grad_y S$ & trajectory momentum: $p_t:=P$ \\
     \hline\multicolumn{4}{c}{Born-Huang nuclear wavefunction $\psi_i(y,t)$} \\\hline
   B & phase: $S_i:=\arg(\psi_i)$ & \multicolumn{2}{c}{momentum: $p_i:=\grad_y S_i$} \\\hline
   
   \multicolumn{4}{c}{transformation  a.u. $\rightarrow$  scaled units} \\
    \hline
      C & $\alpha_{coh}:=M\omega$ (a.u.) &  $\sqrt{\alpha_{coh}}y\rightarrow y$ & 
       $\sqrt{\alpha_{coh}}q_t\rightarrow q_t$ \\
      D & $\alpha_t/\alpha_{coh} \rightarrow \alpha_t$ & $\beta_t/\alpha_{coh}\rightarrow \beta_t$ & $\Bar{p}_t/\sqrt{\alpha_{coh}}\rightarrow \Bar{p}_t $ \\\hline
   \end{tabular}
    \caption{The model parameters and definitions.   Rows A and B define quantities of the nuclear wavefunctions. Rows C and D define transformation  from the atomic units to scaled units. 
   }
    \label{tab:Definitions}
\end{table}
 To streamline the analysis, the nuclear wavepacket  $\psi_1(y,0)$  on the ground electronic surface is taken as the vibrational ground state of 
 $V_{11}=(M\omega^2y^2\!-\!\omega)/2$, its width parameter equal to the coherent value, $\alpha_{coh}$,     $\alpha_{coh}=M\omega$, 
\be \psi_1( {y},t):=
\sqrt{1\!-\!\lambda^2}  \left(\frac{  {\alpha}_{coh}}{\pi}\right) ^{1/4} \exp\left(-\frac{ {\alpha}_{coh} y^2}{2}\right)
\label{eq:psi1}.\ee  In the absence of coupling  the wavepacket $\psi_1$  is time-independent. 

We assume that at $t=0$, the laser pulse creates a   Gaussian wavepacket $\psi_2$ in the excited electronic state with  population of  $\lambda^2$, 
\be \psi_2( {y},t) :=
\lambda \left(\frac{  {\alpha}_t}{\pi}\right) ^{1/4} \exp\left(-\frac{ {\alpha}_t+\imath {\beta}_t}{2}(y-q_t)^2+\imath  \Bar{p}_t(y- {q}_t)+\imath\gamma_t\right) \label{eq:psi2}.
\ee
 For $V_{22}=-ky-\omega/2$, $\psi_2$ depends on time through its parameters as follows:  
\bea  \label{eq:traj}
 \Bar{p}_t & = & \Bar{p}_0+ kt \label{eq:GWPMom}, ~~
 {q}_t  =  q_0+\frac{ \Bar{p}_0}{M}t+ \frac{k}{2M}t^2\\\label{eq:width}
 {\alpha}_t & = & \frac{ {\alpha}_0}{ 1+ \tau^2},~~ {\beta}_t  =  \beta_0-\frac{ \alpha_0 \tau}{1+ \tau^2},~~ \mbox{~where~}\tau:=\frac{\alpha_0t}{M} \\\label{eq:gamma}
  {\gamma}_t & = &\gamma_0+ k {q}_0t+ \frac{t}{M}
  \left(\frac{k^2t^2}{3}+k \Bar{p}_0t+\frac{ \Bar{p}_0^2}{2}\right)+\frac{\omega t\!-\!\arctan{\tau}}{2}.  
\eea
Henceforth, we will use the {\it scaled variables and parameters} of the  wavefunctions,  defined in  Table \ref{tab:Definitions}, without introducing new notations. In the scaled variables $\alpha_{coh}=1$.  

\subsection{Definition of  the XF electronic coefficients}
The XF electronic coefficients $C_i$ ($i=1,2$) in Eq. (\ref{eq:eansatz})  are  obtained by equating the XF wavefunction given by Eqs (\ref{eq:XFWF}) and (\ref{eq:eansatz}) to the usual Born-Huang (BH) expansion:
\be \Psi(x,y,t)=\psi(y,t)\sum_{i=1,2}C_i(y,t)\phi_i(x,y) = \sum_{i=1,2} \psi_i(y,t) \phi_i(x,y) \label{eq:BHExpansion}. \ee
This is equivalent to the procedure of Ref. \cite{XFDensityMatrix}, where the XF solution is derived from the eigenvectors of the density matrix yielding 
\be C_i=\frac{\psi_i}{\psi}. 
\label{eq:Ci}\ee 
The XF coefficients will be analyzed in terms of their real moduli, $|C_i|$,  and phases, $\theta_i$,  
\be C_i=|C_i|\exp(\imath \theta_i). \label{eq:theta}  \ee 
Since $\Phi$ satisfies    the partial normalization condition of  Eq. (\ref{eq:normalization}), 
the XF nuclear probability density is equal to 
\be |\psi|^2=|\psi_1|^2+|\psi_2|^2 \label{eq:XFNWF},  \ee 
and given  that  $\{\phi_i(x,y)\}$  are orthonormal for each $y$,  $|C_i|^2$    are:  
\be |C_i|^2=\frac{|\psi_i|^2}{|\psi_1|^2+|\psi_2|^2} \label{eq:C1}. \ee

From Eq. (\ref{eq:C1}) it is already clear that as the nuclear BH wavepackets diverge,  \\
$\bra |\psi_1| \mid |\psi_2|\ket\rightarrow 0$,    the amplitudes  $|C_i|$  become either  0 or 1. 
As discussed in Ref. \cite{XFSteps} and illustrated in Fig. \ref{fig:Model}(b),   the amplitudes $|C_i|$  behave as  step-functions  near the point of equal probability density associated with the two electronic states.   The step-functions are also present in the phases of the  XF coefficients, $\theta_i$,    defined  as the difference of the phases of  $\psi_i$ (the nuclear wavepackets  in the BH representation of Eq. (\ref{eq:BHExpansion})) and  the phase $S$ of the  nuclear  XF function, $\psi(y,t)$,  $S(y,t):=\arg(\psi(y,t))$.  In {\it one nuclear dimension}  the latter is obtained (up to a constant) as the integral of  its gradient, $P(y,t):=\grad_y 
S(y,t)$, which matches the QT definitions  of  Eqs (\ref{eq:QTWF}) and (\ref{eq:QTMOM}). 
Within the XF framework $P(y,t)$  is uniquely defined 
as the $x$-averaged nuclear momentum of the full wavefunction $\Psi$: 
\be 
P(y,t)= \Im\bigg(\frac{\bra \Psi | \grad_y | \Psi \ket_x}{\bra \Psi|\Psi\ket_x}\bigg)=
\frac{p_1 |\psi_1|^2+p_2|\psi_2|^2}{|\psi_1|^2+|\psi_2|^2} \label{eq:Py}, \ee
where $\{p_i\}$  denote the  momentum of each BH wavepacket, 
\be p_i(y,t):=\Im\left(\frac{\grad_y \psi_i}{\psi_i}\right).\ee
Then, the phases, $\theta_i$, of the XF coefficients $C_i$  are 
\be \theta_i(y,t) = \arg(\psi_i(y,t))-S(y,t).\label{eq:CPhase}\ee 
 
\section{Results and discussion} \label{sec:TheoryAnalysis}  

In this Section we argue that 
the step-function character of the electronic coefficients underlies  the  challenges of the numerical implementation of the XF-type methods.
Unless specified otherwise, the model and the initial wavefunction  parameters    given  in Table \ref{tab:ModelParamsXFPD}  are used in the analysis. 

\begin{table}
    \centering
    \begin{tabular}{c|c|c|c|c|c}
    \hline
    \multicolumn{6}{c}{Model parameters (scaled units)} \\\hline
       $M=1$ & $\omega=1$ & $k=2$ & $V_{1}(0)=-\frac{1}{2}$ & $V_2(0)=-\frac{1}{2}$ & \\\hline
    \multicolumn{5}{c}{Wavefunction parameters (scaled units)} \\\hline
      $\alpha_0=1$ &  $\beta_0=0$ &  $q_0=0$ &  $\Bar{p}_0=0$
      $\gamma_0=0$   & $\lambda=1/\sqrt{3}$ \\\hline
    \end{tabular}
    \caption{Parameters of the model and initial wavefunction   used in the analysis unless stated otherwise.    }
    \label{tab:ModelParamsXFPD}
\end{table}

 The time-evolution equations are obtained  from  the TDSE (\ref{eq:eTDSE}) for the XF coefficients in polar form  of Eq. (\ref{eq:theta}).    Using the derivative operators, 
 defined in  Eqs  (\ref{eq:D1}, \ref{eq:D2}),  
 the formal EOMs in the stationary frame are: 
\bea \frac{\pd |C_i|}{\pd t}&=&~\Im\left(e^{-\imath \theta_i} ((\hat{D}_1\!+\!\hat{D}_2) C_i)\right)-V_i|C_i| \label{eq:pdtA} \\
 \frac{\pd \theta_i}{\pd t}&=&-\Re\left(C_i^{-1}((\hat{D}_2\!+\!\hat{D}_2)C_i)\right)+V_r-H^{BO}_{ii}. \label{eq:C_formal}
 \eea  
In the moving frame  defined by a QT,  $\dot{y}_t=p_t/M$ (Eq. (\ref{eq:LagFrame})),  the counterpart to Eq. (\ref{eq:pdtA})  becomes  
\bea \frac{d|C_i|}{dt}&=&\frac{\pd|C_i|}{\pd t}+\frac{p_t}{M}\grad_y|C_i|, \label{eq:dtA} 
\eea
 with  the last  RHS term  canceling part of the  term containing $\hat{D}_1C_i$, and  likewise for $d\theta_i/dt$.  However, for simplicity  we will use the explicit form of $|C_1(t)|$  from  Eqs (\ref{eq:psi1}, \ref{eq:psi2}) and (\ref{eq:C1}). As follows from Eq. (\ref{eq:C1}) time-dependence of  $|C_2(t)|$ is complementary to that of  $|C_1(t)|$.      As noted previously  in the one-dimensional case, $V_r$ can be chosen such that  $V_i=0$.   



We will analyze the behavior of $C_1(y,t)$ near the point  of  the largest change in the populations, i.e  near the point of equal density between the centers of $\psi_1$ and $\psi_2$,  which will henceforth be referred to as the median, $y_{m}(t)$
\be |\psi_1(y,t)|_{y=y_m}=|\psi_2(y,t)|_{y=y_m}. \label{eq:median}\ee 
 The initial wavefunction $\psi_2$  is defined as $\alpha_0=1, \beta_0=0, q_0=0, \bar{p}_0=0$ (Table \ref{tab:ModelParamsXFPD}).  Equations (\ref{eq:pdtA}) and (\ref{eq:dtA}) are  examined   assuming the 'frozen'  wavepacket approximation for $\psi_2$,    $\alpha_t= 1$, or in other words  in the limit of small $\tau$ in Eqs (\ref{eq:width}).  This approximation  removes the spreading of the QTs, which  simplifies the analysis of the electronic coefficient behavior; while preserving the essential features of the dynamics as shown next. 
 
 The QT positions, $y_t$, obtained from the evolution of the  exact and  frozen  wavepackets  are  
  displayed  in Fig. \ref{fig:XF Traj and Dynamics}(a,c).  The centers of $\psi_2$ and $\psi_1$ (the latter equal to zero for all $t$), and $y_m$ are shown  as red dash and as green dot-dash, respectively.  Comparing these two panels, we see that  $y_m$  moves at a faster rate  in the case of  frozen $\psi_2$  ($\alpha_t=1$), but follows the same trend as the exact $y_m$.    In the frozen  case $y_m$ is equal to 
 \be y_{m}=\frac{q_t}{2}+\frac{\ln(\eta)}{2q_t} \label{eq:FWPmedian},   \ee  
 where $\eta$ denotes the ratio of the full populations of the two  electronic states,  
 \be \eta:=\frac{1-\lambda^2}{\lambda^2}. \label{eq:eta}\ee
 The moduli of the corresponding  electronic coefficients, $|C_1(y_t)|$, are displayed in Fig.  \ref{fig:XF Traj and Dynamics}(b,d).  We observe  that  the frozen wavepacket   approximation does not change  the qualitative dependence of $|C_1|$, and that the largest changes in the state  populations of  QTs, $|C_1(y_t)|^2$,  indeed,  occur  near $y_t=y_m$, marked with the open circles.

\begin{figure}
    \centering
    \begin{tabular}{l|l}
    \hspace{2cm} {\large  \it trajectories} & \hspace{2cm} {\large \it populations} \\\hline
  {\large (a) \texttt{ exact}}   &  {\large (b) \texttt{exact}}  \\  \includegraphics[width=0.4\textwidth]{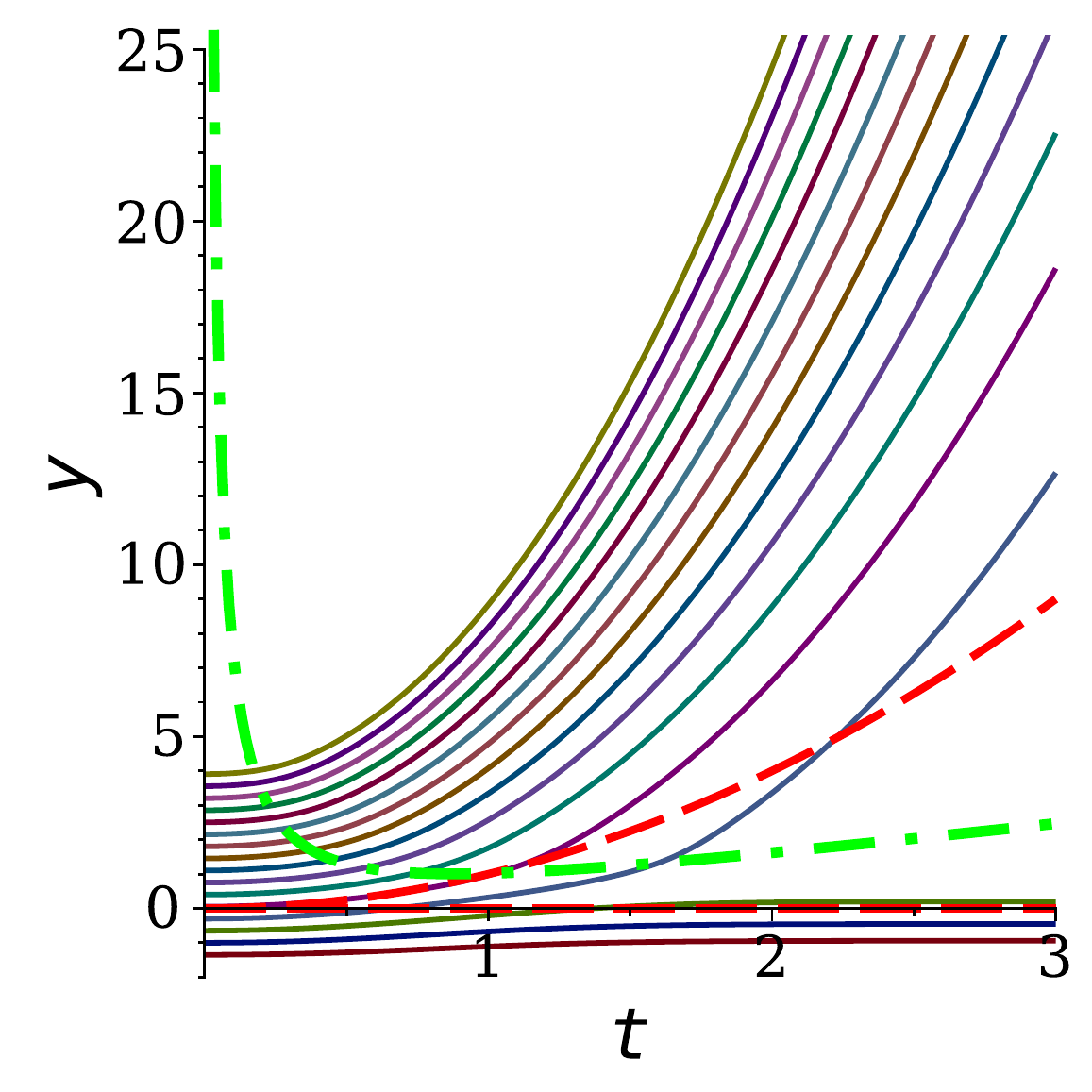}    & \includegraphics[width=0.4\textwidth]{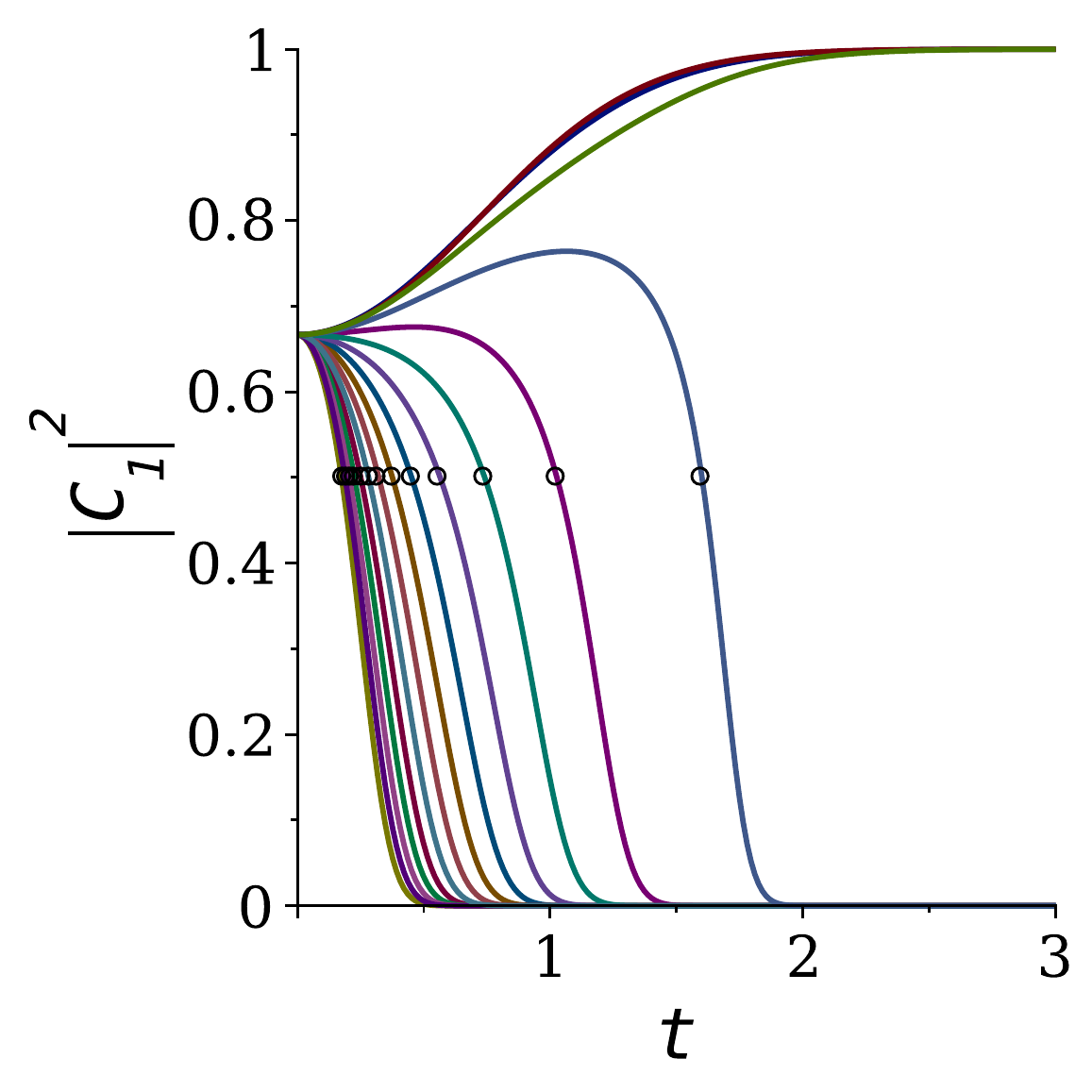}  
     \\\hline  {\large (c)   \texttt{frozen}} &   {\large (d)   \texttt{frozen}}  \\
     \includegraphics[width=0.4\textwidth]{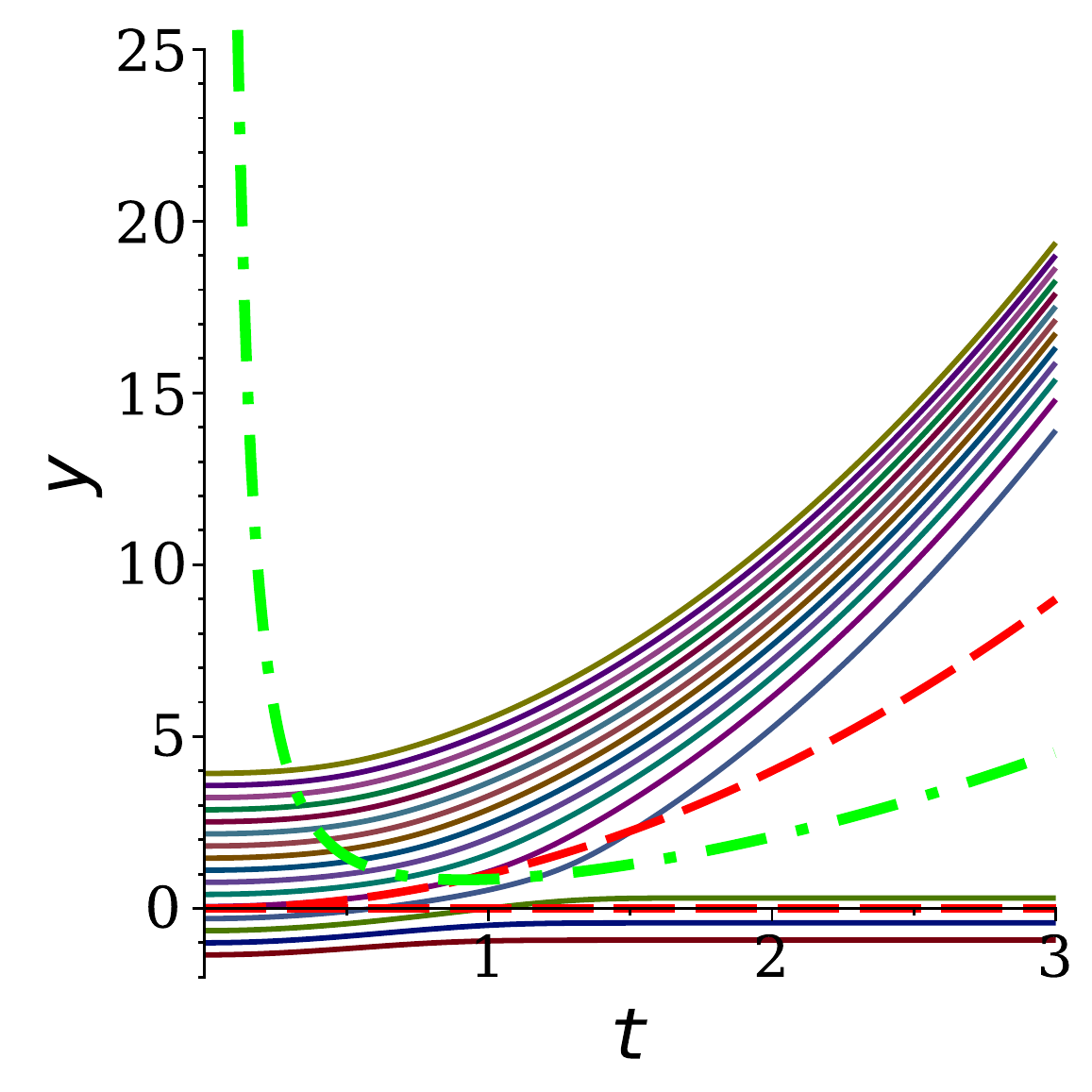}    &\includegraphics[width=0.4\textwidth]{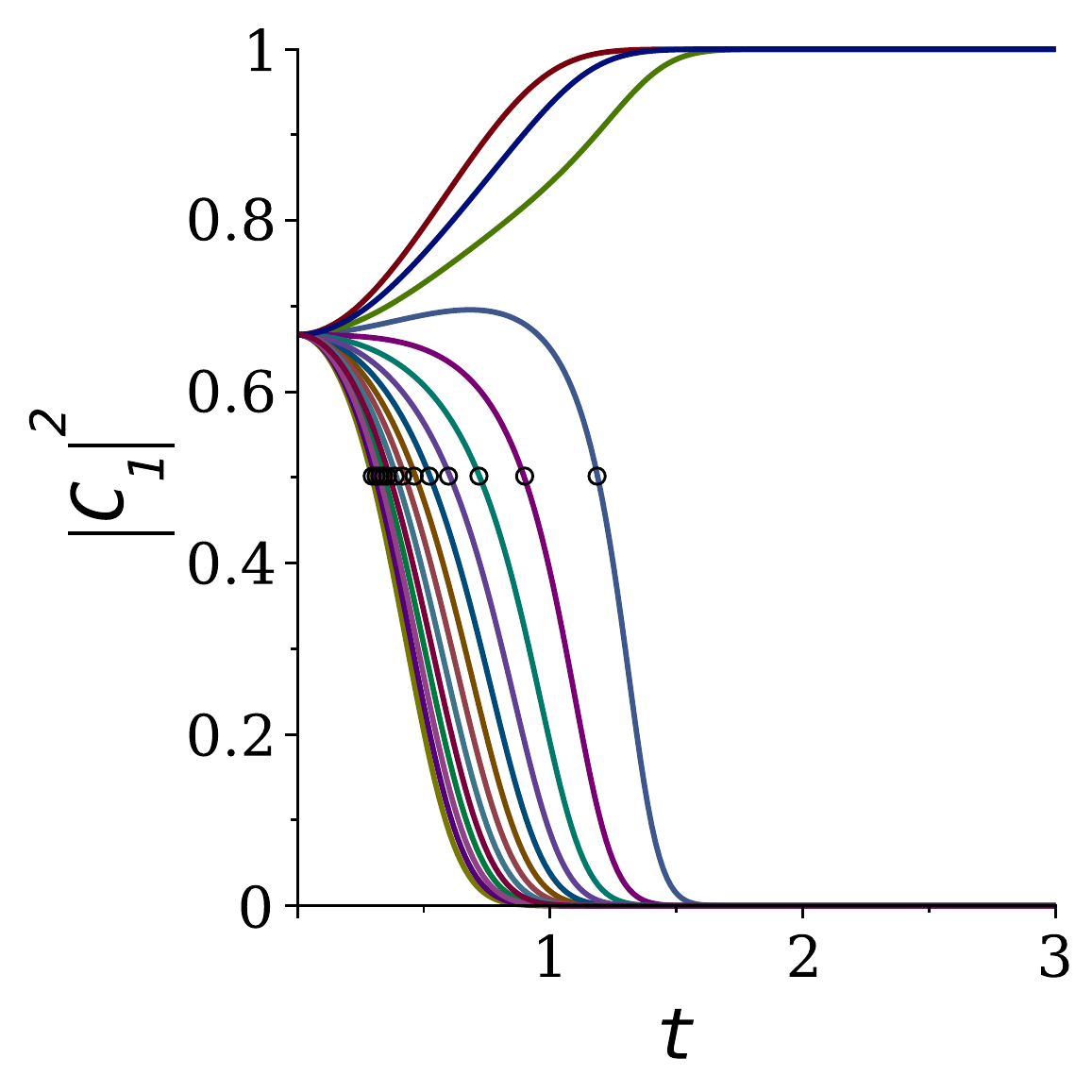}  
    \end{tabular}
    \caption{The quantum trajectories  and associated  electronic populations for the photodissociation model.   Panels (a) and  (c)  show the QTs (solid lines,  Eq. (\ref{eq:rhoc})) for the exact and frozen wavepackets, respectively.  The red dashed lines are the centers of the ground and excited state Gaussians (note the ground state center remains at zero for all time). The green dot dash line marks the position of equal density,  $y_m(t)$. Panels (b) and (d)  show the populations of the electronic ground state along each trajectory, with the colors of the solid lines corresponding to trajectories plotted in (a) and (c) correspondingly. The black open circles indicate the point where the dissociating trajectories  cross $y_m(t)$.  }
    \label{fig:XF Traj and Dynamics}
\end{figure}

To analyze the effect of the moving frame on the behavior of $C_i$ let us  compare Eqs (\ref{eq:pdtA}) and (\ref{eq:dtA}). 
The QTs are defined by an integral of the nuclear probability density, Eq. (\ref{eq:rhoc}), which yields: 
\be \rho_c=\frac{1}{2}\big(1+\text{erf}(y_t)\big)+\frac{\lambda^2}{2}\big(\text{erf}(y_t-q_t)-\text{erf}(y_t)\big) \label{eq:QTeq}. \ee
The velocities of the QTs are then:
\be \Dot{y}_t=\frac{\Dot{q_t}}{1+\eta\sigma}  = \frac{\bar{p}_t}{M(1+\eta\sigma)} \label{eq:QTVelocSimpler}, \ee 
where $\sigma$ denotes the probability density ratio of the BH nuclear wavepackets, 
\be \sigma:=\frac{\exp(-y_t^2)}{\exp(-(y_t-q_t)^2)}.\ee

The XF coefficient  $C_1$ is known up to a phase,  
\be C_1(y,t) = \frac{\sqrt{1-\lambda^2}e^{-y^2/2+\imath \theta_1(y,t)}}{\left((1-\lambda^2)e^{-y^2}+\lambda^2 e^{-(y-q_t)^2}\right)^{1/2}},\ee
which is not needed for  the time-derivatives of  $|C_1|$. 
Using $\dot{q}_t=\bar{p}_t/M$,  the Eulerian EOM for $|C_1|$ is 
\be  \frac{\pd |C_1|}{\pd t}  = \frac{\bar{p}_t}{M}\frac{\sqrt{\eta\sigma}}{(1+\eta\sigma)^{3/2}} (q_t-y).  \label{eq:analyticpdapdt}\ee 
In the moving QT frame  defined by  $\dot{y}_t$ of Eq. (\ref{eq:QTVelocSimpler}), 
the time-dependence of $|C_1|$  is defined by the following equation:  
\be \frac{d |C_1|}{dt}=   \frac{\bar{p}_t}{M}\frac{\sqrt{\eta\sigma}}{(1+\eta\sigma)^{3/2}} \left(\frac{\eta\sigma}{1+\eta\sigma}q_t-y_t\right).   \label{eq:analyticdadt}\ee
 Substitution of  $y_m$ (Eq. (\ref{eq:FWPmedian})) into Eq.  (\ref{eq:analyticpdapdt})   
  gives   \be \left.\frac{\pd |C_1|}{\pd t}\right|_{y=y_m} =  \frac{\Bar{p}_t q_t}{\sqrt{32}M} - \frac{\Bar{p}_t \ln(\eta
)}{\sqrt{32}M q_t}, \label{eq:eul_med}\ee  
while  in the QT frame  Eq. (\ref{eq:analyticdadt}) yields: 
\be \left.\frac{d |C_1|}{d t}\right|_{y_t=y_m}  =  - \frac{\Bar{p}_t \ln(\eta)}{\sqrt{32}M q_t}. \label{eq:lagr_med}   \ee
In Eqs (\ref{eq:eul_med}) and (\ref{eq:lagr_med}) the term $\sim 1/q_t$  goes to zero for large $q_t$ (long times), and  for $\eta=1$  the latter is equal to  zero at all times.  

Formally, the behavior of    the  time-derivatives of $|C_1(y,t)|$  are  defined by the   differential operators $(\hat{D}_1+\hat{D}_2)C_1$ in Eq. (\ref{eq:C_formal}). Thus,  if $d|C_1|/dt$  
  in the QT framework  is small, then we could expect the   combined action of the derivatives in the RHS of Eq. (\ref{eq:C_formal})  to be well-behaved numerically despite the step-function character of $|C_1(y,t)| $ at long times. To analyze this numerical aspect we examine the extrema of Eqs (\ref{eq:analyticpdapdt}) and (\ref{eq:analyticdadt})  by  employing the following   approximation: noting that the exponential dependence on $y_t$ in $\sigma$ is larger than the linear dependence of other terms,  these two equations  are differentiated with respect to $\sigma$ and the result is  solved for $\sigma$. With that the positions of the extrema,  denoted $y_\delta$, satisfy the following conditions at $y=y_\delta$ or $y_t=y_\delta$ for the Euler and Lagrange frames, respectively:  
  \be \sigma_{Euler}=\frac{1}{2\eta} \label{eq:sigmaeul} \ee
  \be \sigma_{Lagrange}=\frac{y_\delta+3q_t - \sqrt{9q_t^2-2q_t y_\delta +9y_\delta^2}}{4(q_t-y_\delta)\eta}  \label{eq:sigmalag}. \ee  
  As demonstrated in Fig. \ref{fig:dcdt}, the extrema occur close to $y_m$. Thus,  we  represent $y_\delta$  through its displacement, $\delta$,  from $y_m$,   
  \be y_{\delta}=y_{m}+\delta \label{eq:ydelta}, \ee and solve  Eqs (\ref{eq:sigmaeul})  and (\ref{eq:sigmalag}) for  $\delta$ taking the natural logarithm of the equations (Taylor-expanding to the first order in $\delta$ at $\delta=0$ and in $1/q_t$ at $q_t=\infty$  in the later case).  
This yields the following analytic estimates for the positions of the extrema in the Euler and Lagrange frames of reference, respectively:  \be \delta_{Euler}=\frac{\ln(2)}{2q_t} \label{eq:deltaeul} \ee
\be \delta_{Lagrange}=-\frac{1}{2q_t}\ln \left(\frac{7}{4} \pm \frac{\sqrt{41}}{4}\right) +O(q_t^{-3}) \label{eq:deltalag}. \ee
Note that  
the Lagrangian solution 
has two solutions corresponding to the extrema to the left and right of $y_m$ seen in Fig.  \ref{fig:dcdt}. In both frames the extrema positions converge to $y_m$ as $q_t\rightarrow\infty$. The analytic expressions for both frames converge as $O(q_t^{-3})$ to the numerically exact solutions shown in Fig.  \ref{fig:dcdtExtrema}. 
 The extremal values of the time-derivatives of $|C_1|$ are:  
\be \frac{\pd |C_1(y_\delta)|}{\pd t}=\frac{\sqrt{3}\bar{p}_tq_t}{9M} + O(q_t^{-1}) \label{eq:Maxeul} \ee
\be \frac{d |C_1(y_\delta)|}{d t}=\pm\frac{2(\sqrt{41}\pm3)\sqrt{7\pm\sqrt{41}}\bar{p}_tq_t}{M(11\pm\sqrt{41})^{\frac{5}{2}}} + O(q_t^{-1}) \label{eq:Maxlag} \ee 
Disappointingly, the amplitude of the  derivative extrema  in {\em both} frames grows with  time as  $q_t\Bar{p}_t$,  which for this model means as $t^3\propto q_t^{\frac{3}{2}}$,  which matches the behavior of the accurate numerical amplitudes  displayed   in Fig. \ref{fig:dcdtExtrema}(b).  Consistent with  Figs \ref{fig:dcdt} and \ref{fig:dcdtExtrema},  the constant prefactor in front of the Lagrangian extrema is smaller (by a factor of 0.6 in our model),   which may  slightly alleviate the numerical difficulties of applying  the derivative operators $\hat{D}_1+\hat{D}_2$ to  the XF coefficients  at long times (or more generally for large $q_t\Bar{p}_t$). 

\begin{figure}
    \centering
    \begin{tabular}{ll}
       {\large (a)}   &~~~ {\large (b)}   \\
      \includegraphics[width=0.45\textwidth]{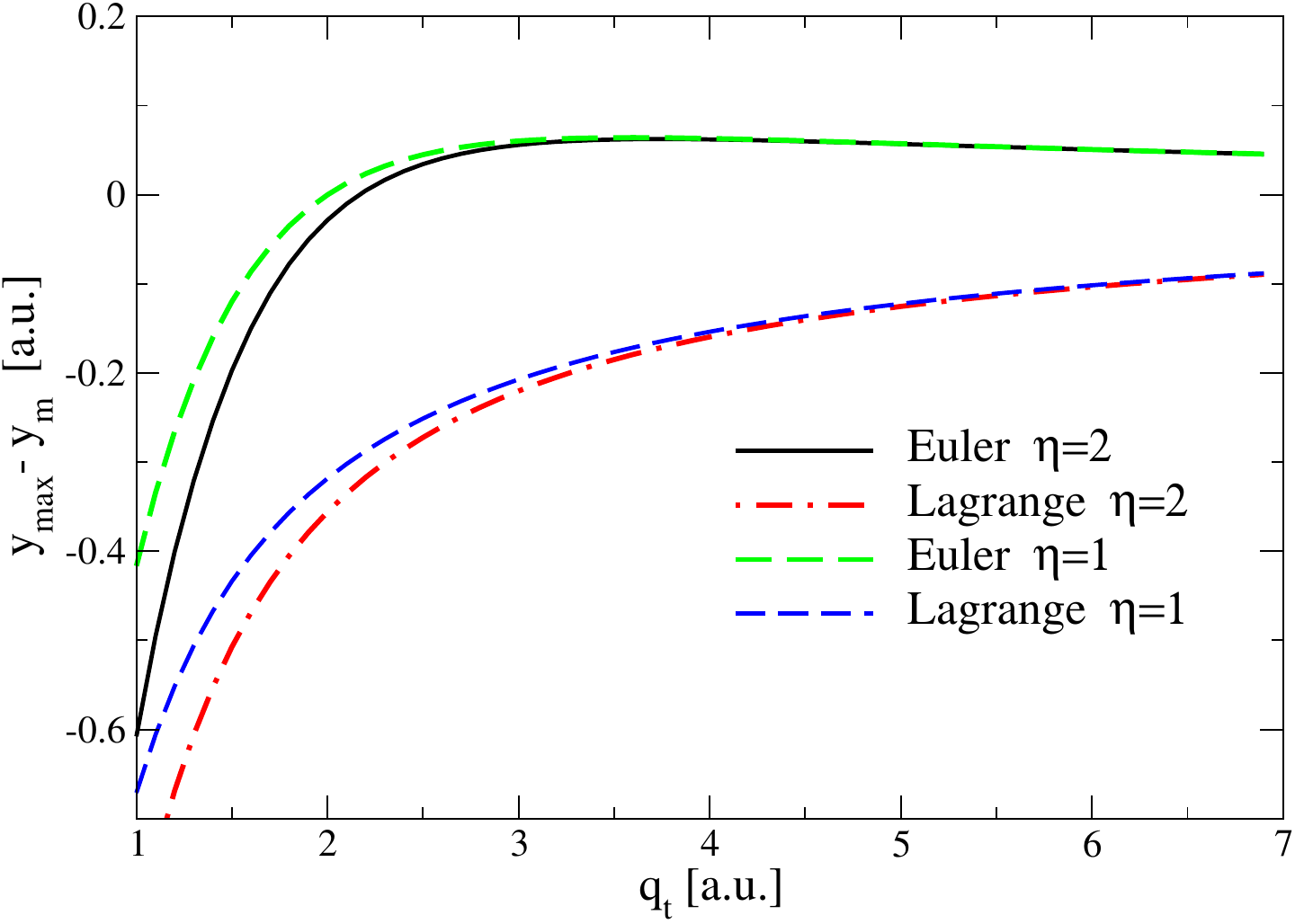}   & ~~~ \includegraphics[width=0.45\textwidth]{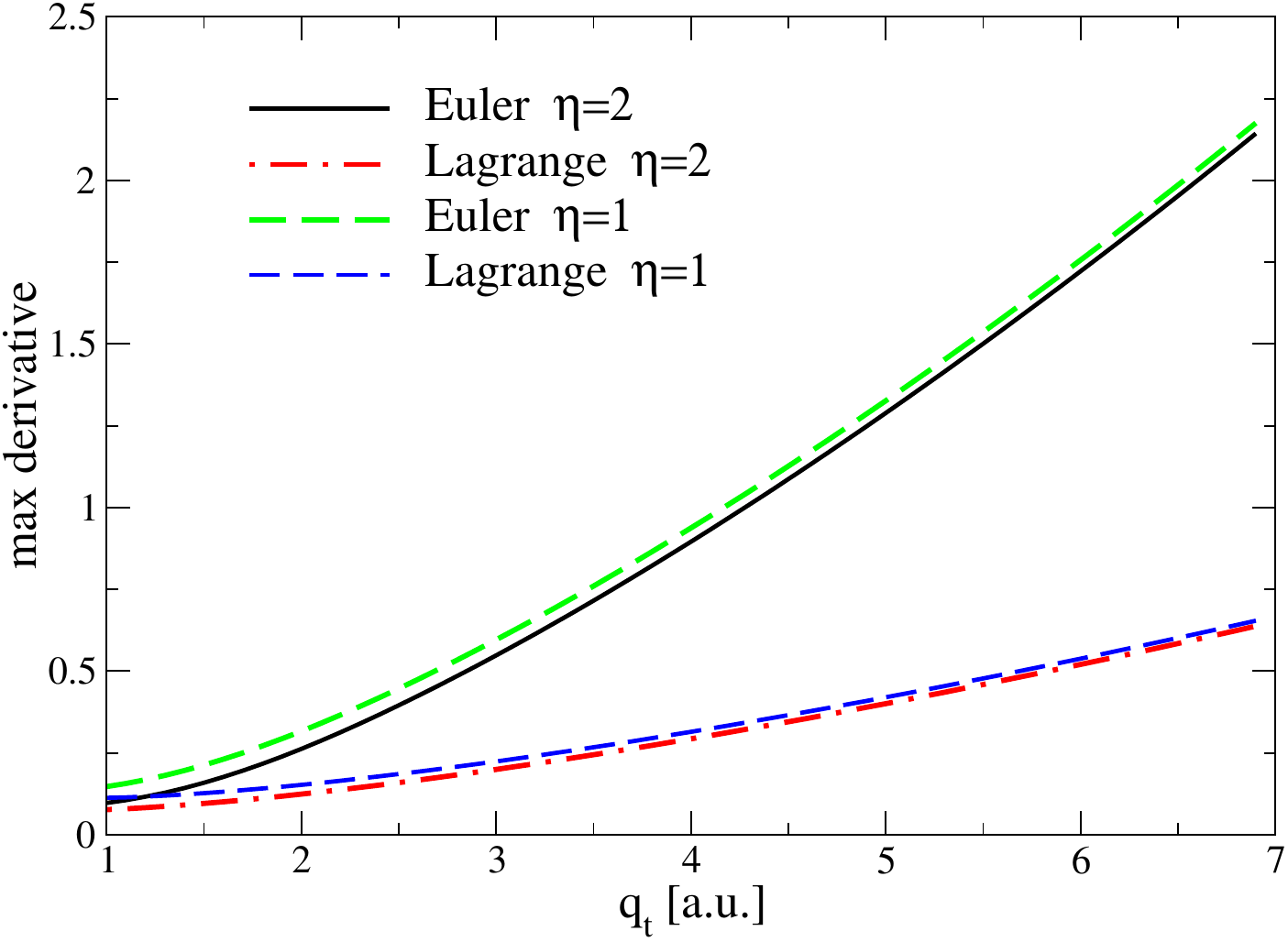}
    \end{tabular} 
    \caption{The maximum time-derivative of $|C_1|$ in the Eulerian and Lagrangian frames of  reference computed for $\eta=1$ and $\eta=2$ using   $M=100$ and $k=20$ a.u. (a) Position with respect to the median,  $y_{max}-y_m$, and (b) the largest  amplitude of the extrema  of  the time-derivative,  $\pd|C_1|/\pd t$ and of $d|C_1|/dt$ as functions of $q_t$ for the Eulerian and Lagrangian frames of reference, respectively.     }
    \label{fig:dcdtExtrema}
\end{figure}

\begin{figure}
    \centering
    \begin{tabular}{lll}
\hline
 $\eta=1/2$ & $\eta=1$ & $\eta=2$ \\
 \hline \\ \multicolumn{3}{c}{
\texttt{Eulerian frame of reference:}  ${\pd |C_1|}/{\pd t}$} \\
(a) & (b)  & (c)\\
 \includegraphics[width=0.32\textwidth]{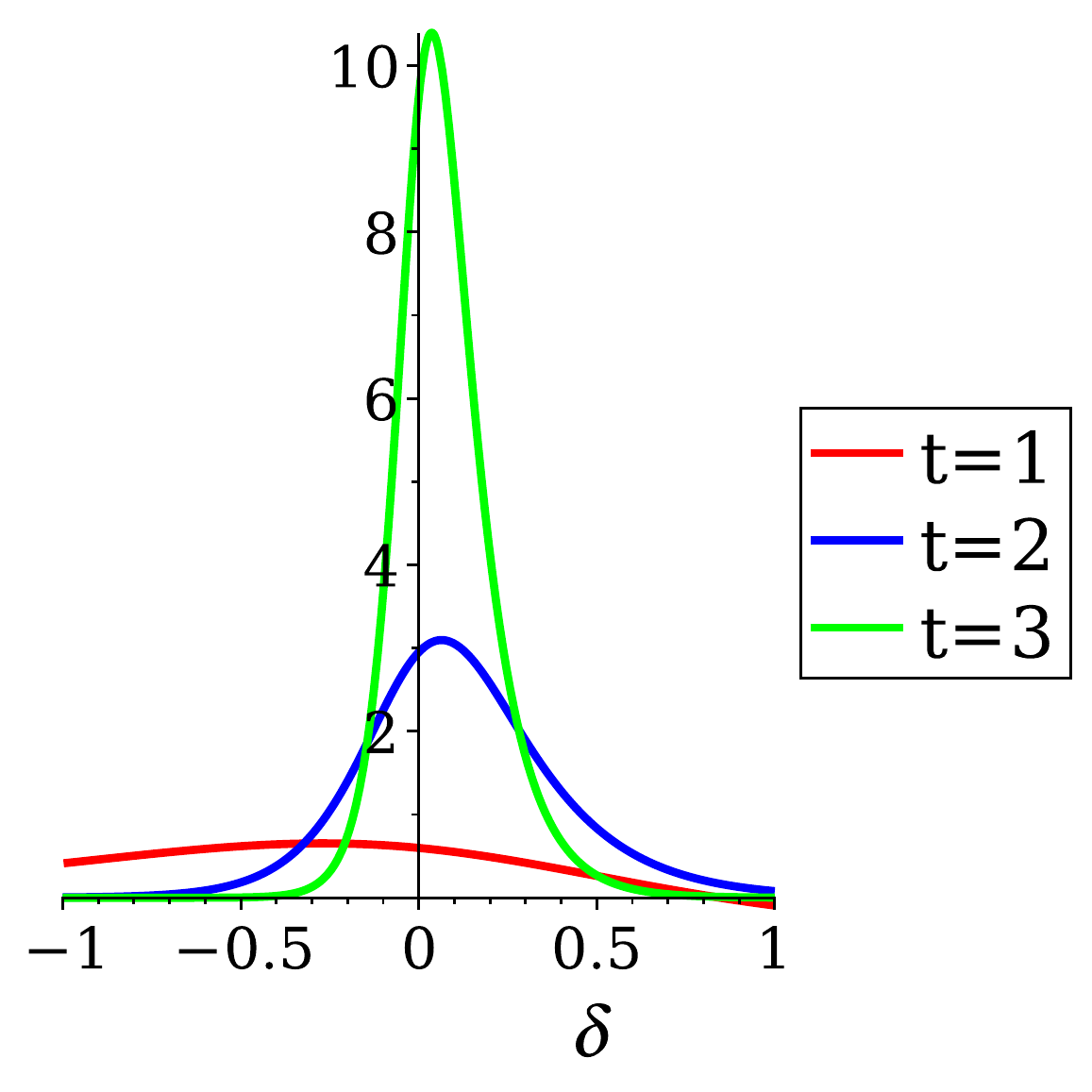} & \includegraphics[width=0.32\textwidth]{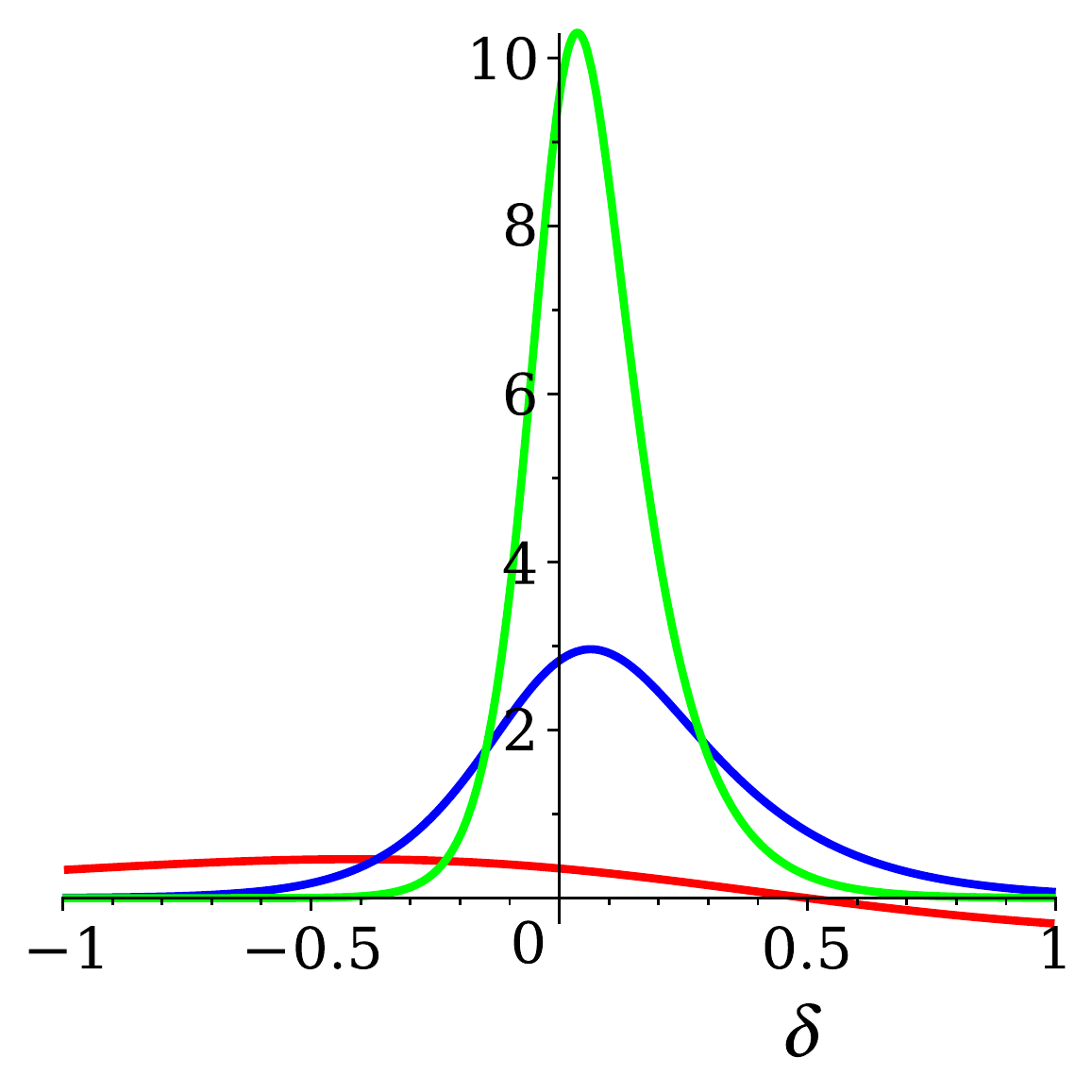} & \includegraphics[width=0.32\textwidth]{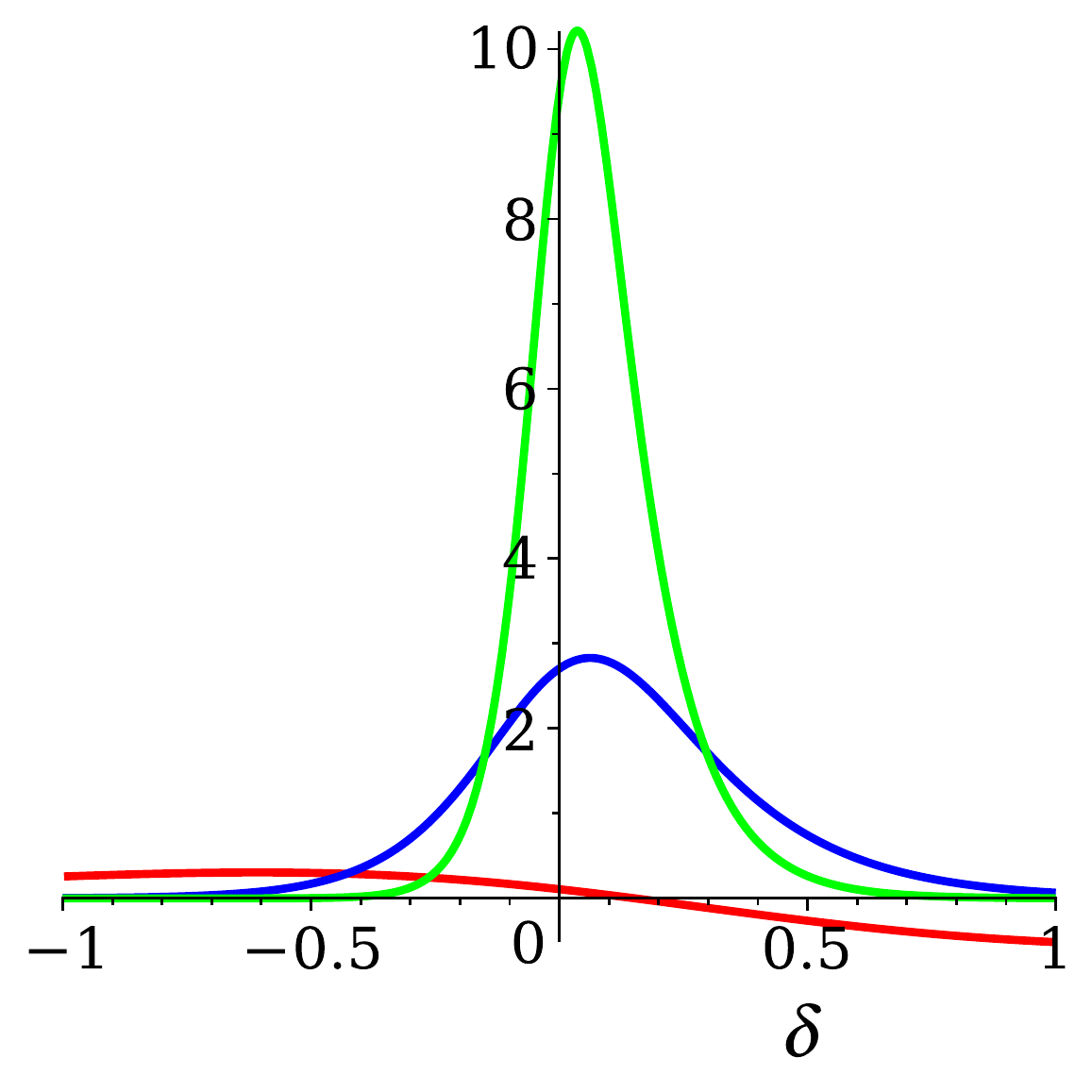} \\
 \multicolumn{3}{c}{\texttt{Lagrangian frame of reference:} ${d |C_1|}/{d t}$} \\
 (d) & (e) & (f) \\
 \includegraphics[width=0.32\textwidth]{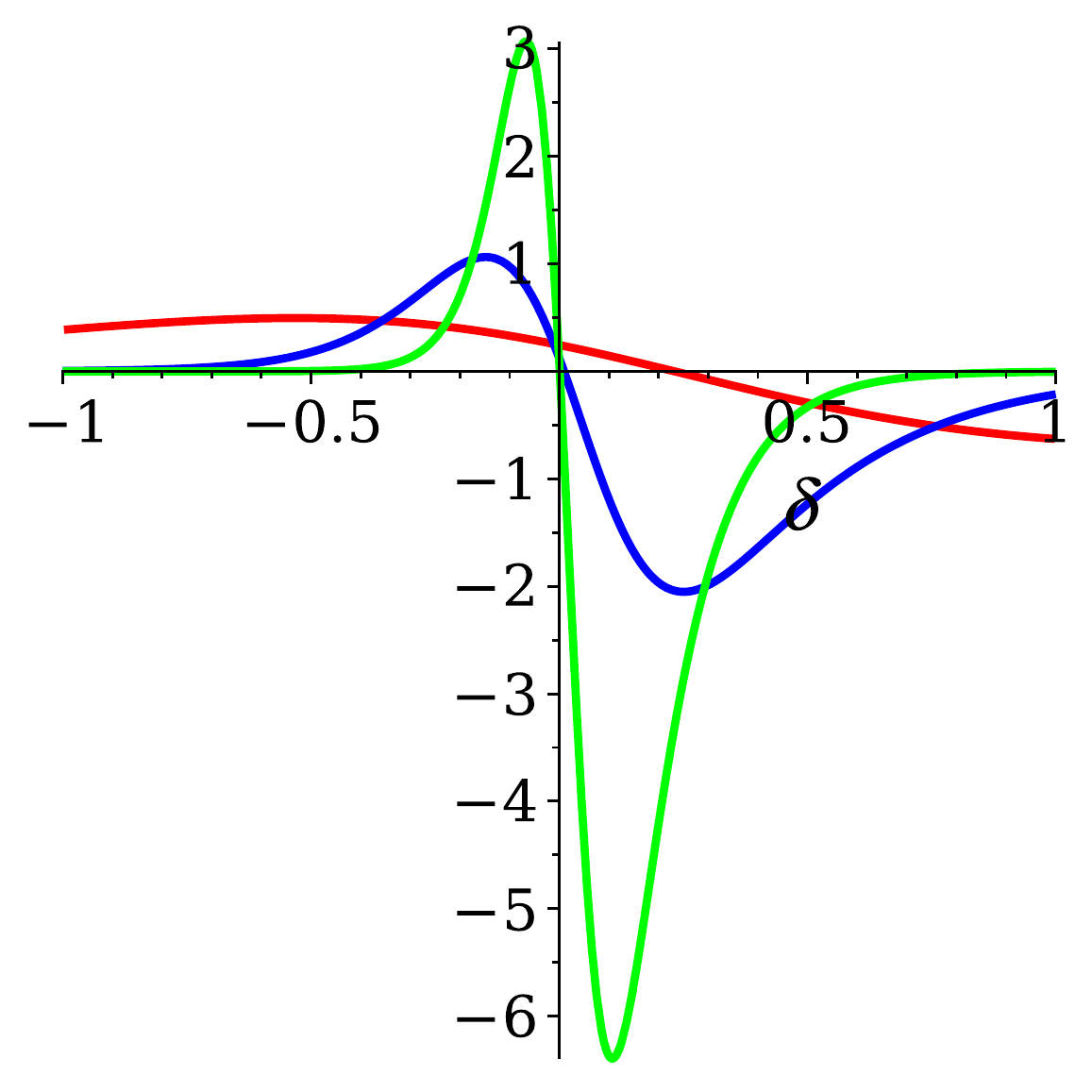} & \includegraphics[width=0.32\textwidth]{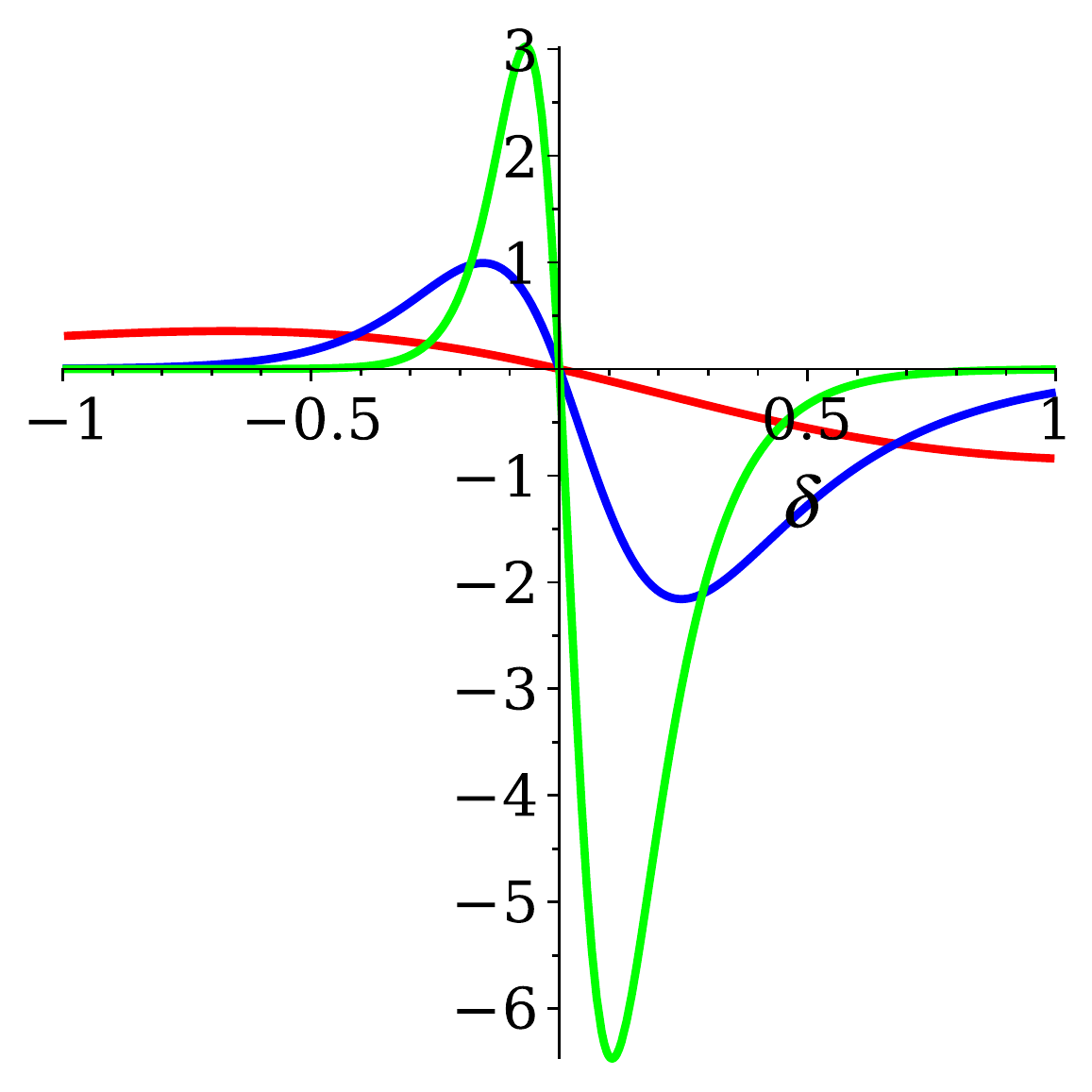} & \includegraphics[width=0.32\textwidth]{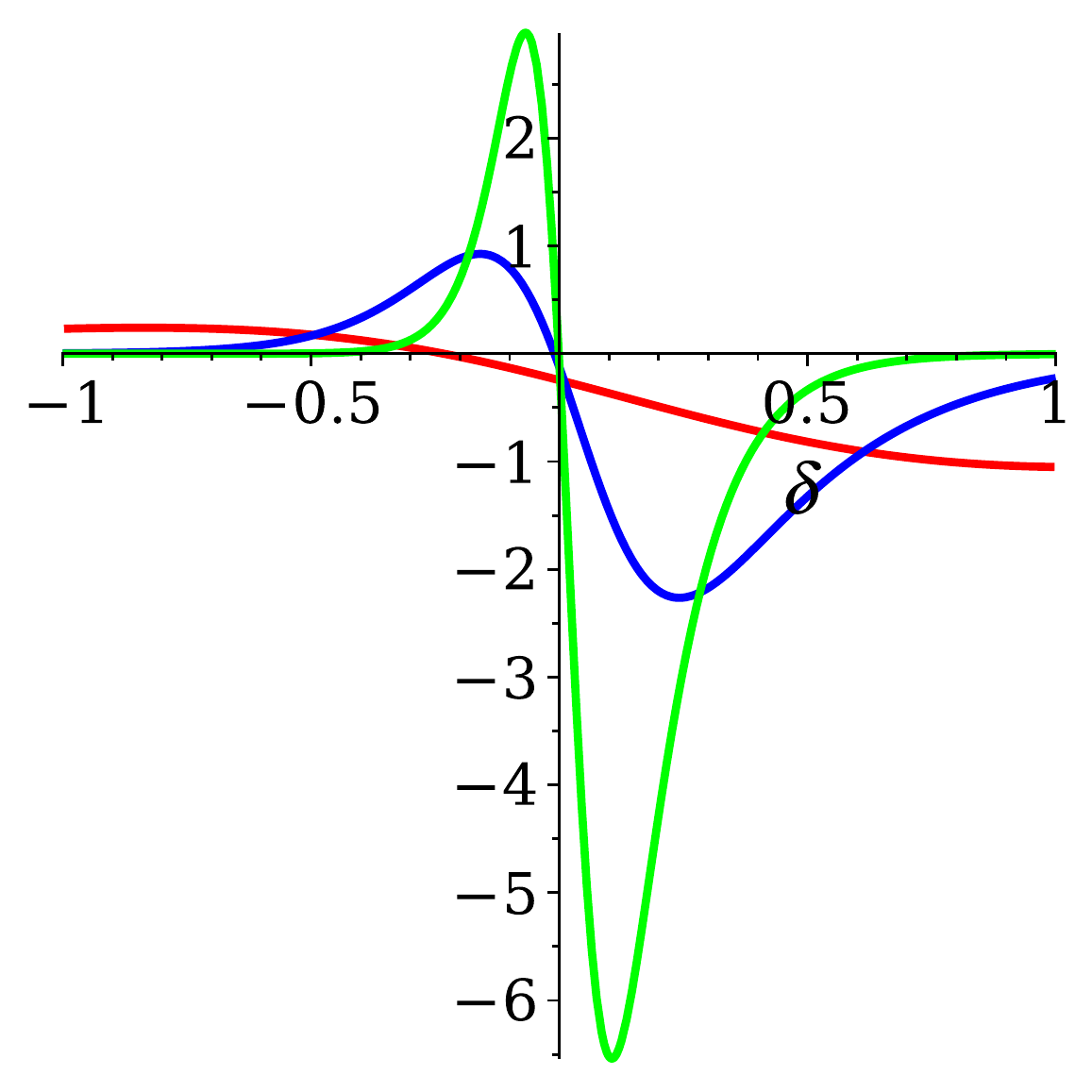}

\end{tabular}
    \caption{Time-derivative of the ground state coefficient amplitude in (a-c) the stationary Eulerian and (d-f) moving  Lagrangian frames with $\delta$ indicating the relative position to the median point. For the moving frame, in the $\eta=1$ case, the point of 0 amplitude change is always at the median; otherwise its relative position changes. The size of the derivatives increases as $q_t\Bar{p}_t$, or  cubically in time as given by Eqs (\ref{eq:analyticpdapdt}) and (\ref{eq:analyticdadt}).
     }
     \label{fig:dcdt}
\end{figure}

To check if the nearly singular derivatives associated with the XF coefficients  discussed above are specific to the BO eigenstate representation of the electronic wavefunction  we also  considered   representation of the molecular wavefunction  in a localized, or "atomic orbital",   basis, 
$\{\chi_L,\chi_R\}$, 
\be \Phi(x,y,y):=C_L(y,t) \chi_L(x,y)+C_R(y,t) \chi_R(x,y). \label{eq:LocalizedPhi} \ee
The subscripts $L$ and $R$ refer the functions associated with the left and right 'atoms',  and $\chi_{R(L)}$  are the atomic orbitals. Assuming  that the atomic orbitals are orthogonal, we represent the BH wavefunctions as $\chi_R\pm\chi_L$,  and obtain a similar to Eqs (\ref{eq:BHExpansion}) factorization: $|\psi|^2=|\psi_L|^2+|\psi_R|^2$,  $ |C_{R(L)}|^2=|\psi_{R(L)}|^2/|\psi|^2$.  
The initial conditions are the same as in Table \ref{tab:ModelParamsXFPD}. Initial $\psi_L$ and $\psi_R$ are obtained from the unitary transformation ($45^\circ$ rotation) of $\psi_1$ and $\psi_2$ which are used to obtain $C_L$ and $C_R$. 
The density snapshots, obtained numerically,  are shown    in Fig. \ref{fig:AOs} for $t=[0,0.64,1.28,1.92,2.56,3.20]$ a.u. 
The XF nuclear density and the BH ground state density, $|\psi_1|^2$ in the molecular basis are shown as black and red solid lines.  
The  atomic basis  amplitudes, $|C_L|^2$ and $|C_R|^2$  are  shown as green and blue dashes. 
For $t>0$ the amplitudes  $|C_L|^2$ and $|C_R|^2$ exhibit an  interference pattern which becomes localized at the median point rather than  the step in the XF amplitudes, $|C_1|^2$ and $|C_2|^2$, for the original BH state basis. As the separation increases, the frequency of these oscillations increase {in both time and space} suggesting  that the numerical challenges will persist in the atomic basis as well, and are inherent to the XF representation if the dynamics is characterized by the  diverging wavepackets. 

\begin{figure}
   \centering
   \begin{tabular}{lll}
   (a) t=0 a.u. & (b) t=0.64 a.u. & (c) t=1.28 a.u.\\
  \includegraphics[width=0.32\textwidth]{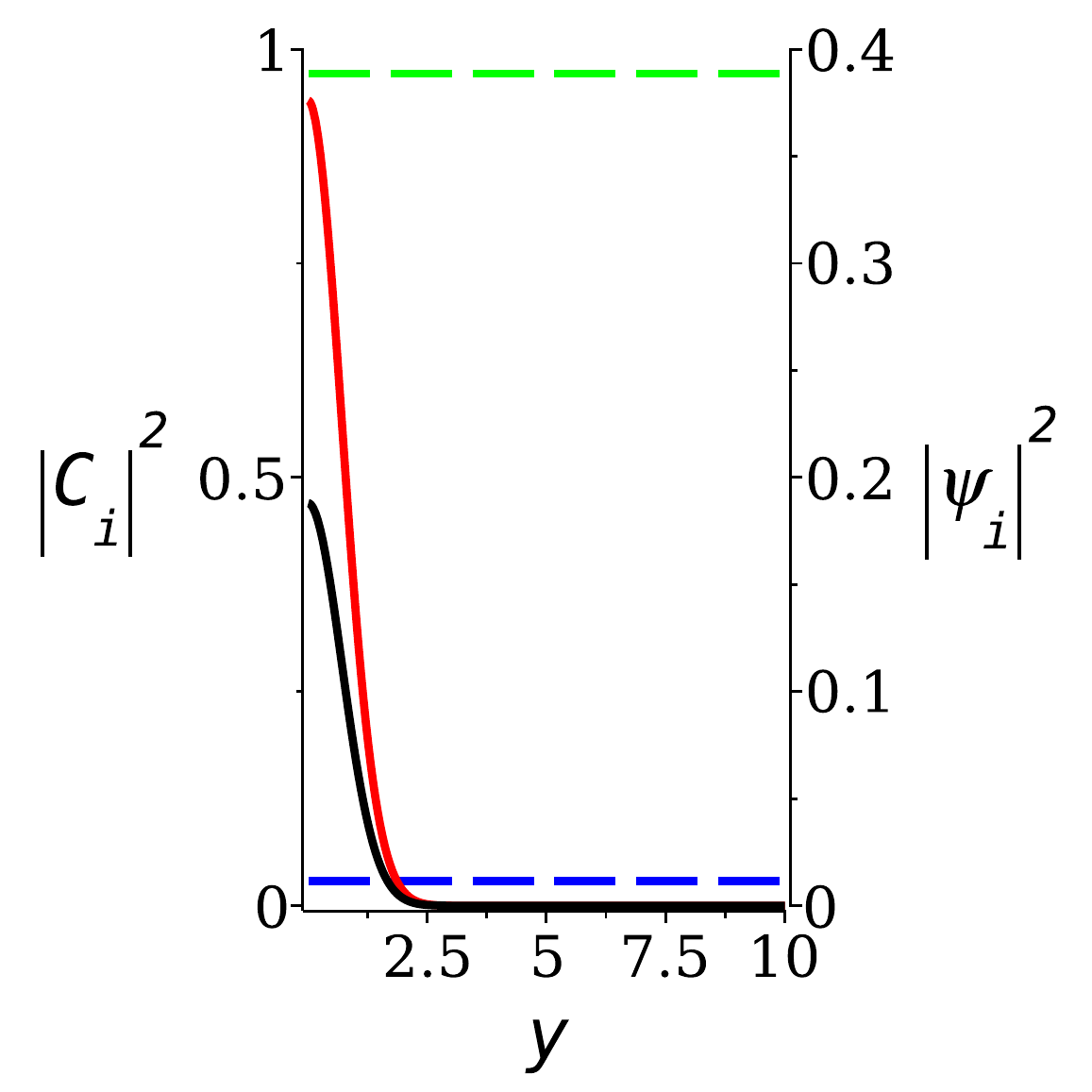}& \includegraphics[width=0.32\textwidth]{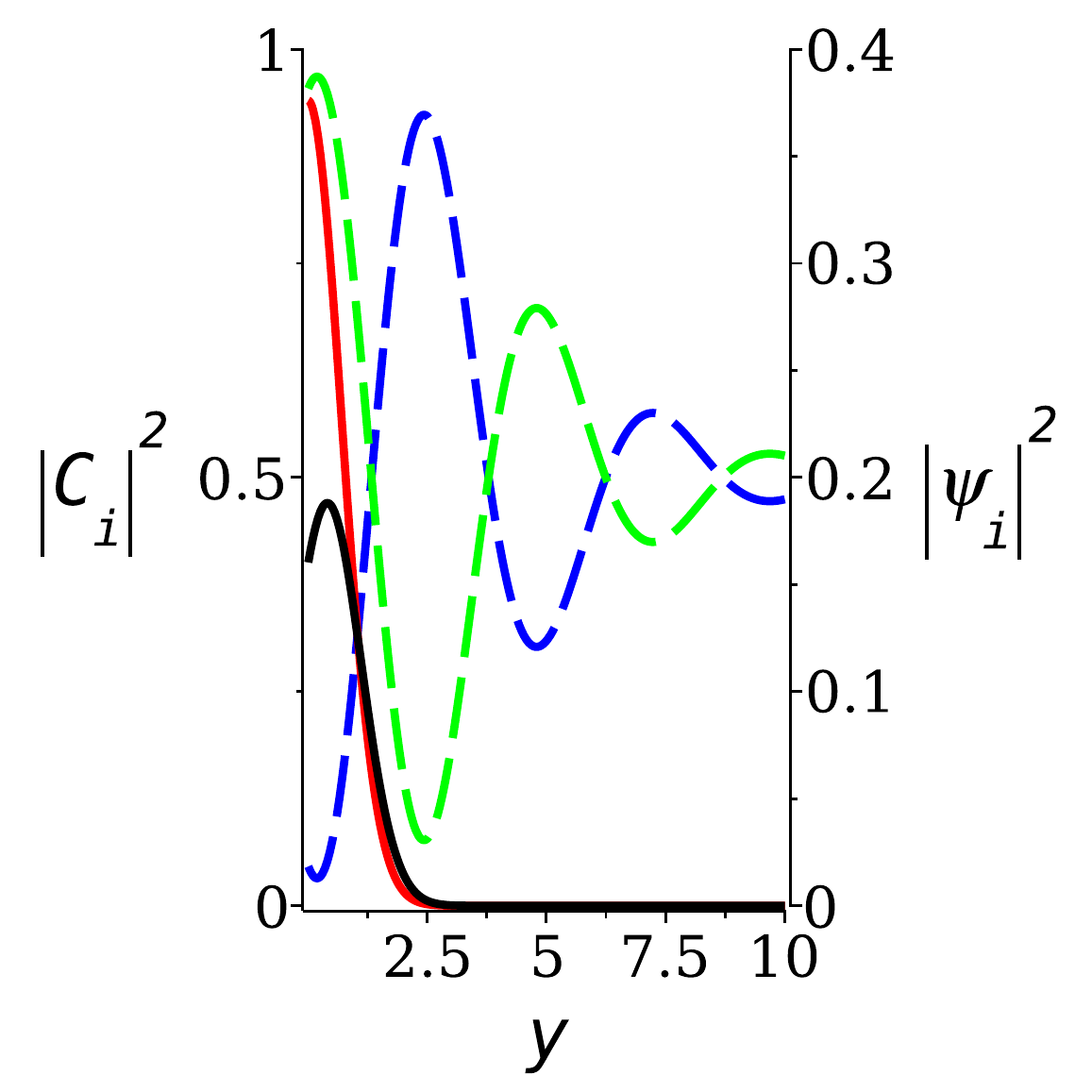} & \includegraphics[width=0.32\textwidth]{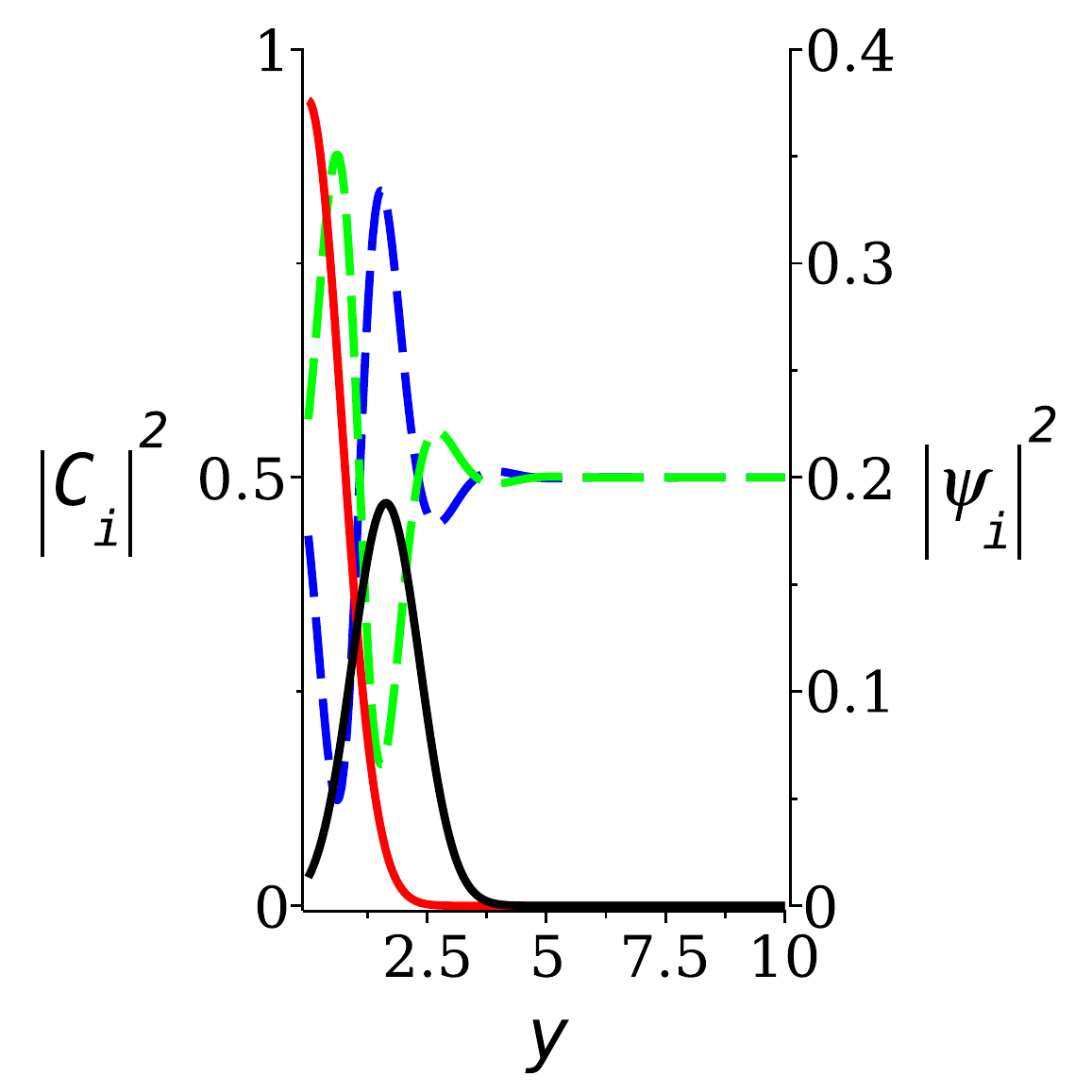} \\
  (d) t=1.92 a.u. & (e) t=2.56 a.u. & (f) t=3.2 a.u. \\
  \includegraphics[width=0.32\textwidth]{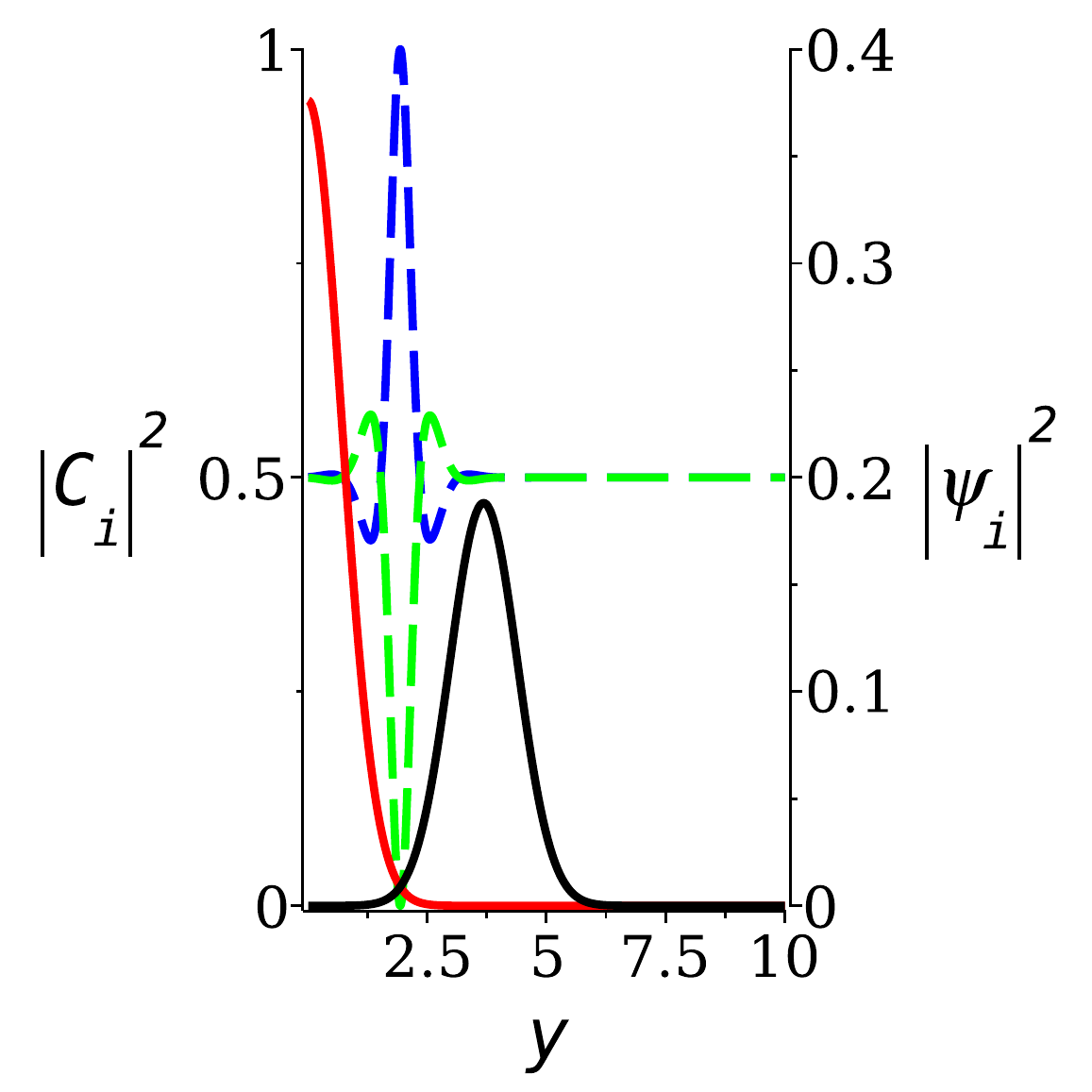} & \includegraphics[width=0.32\textwidth]{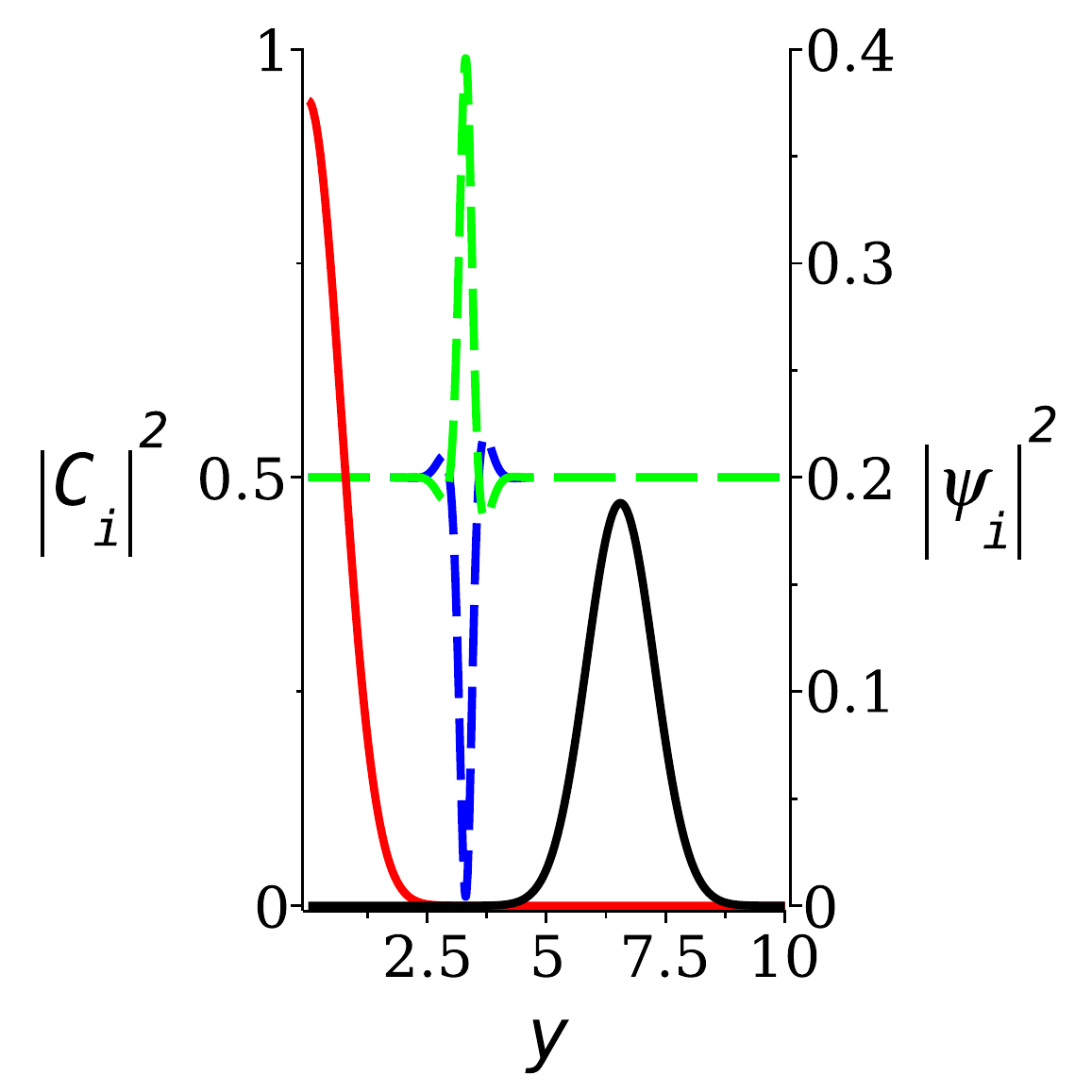} & \includegraphics[width=0.32\textwidth]{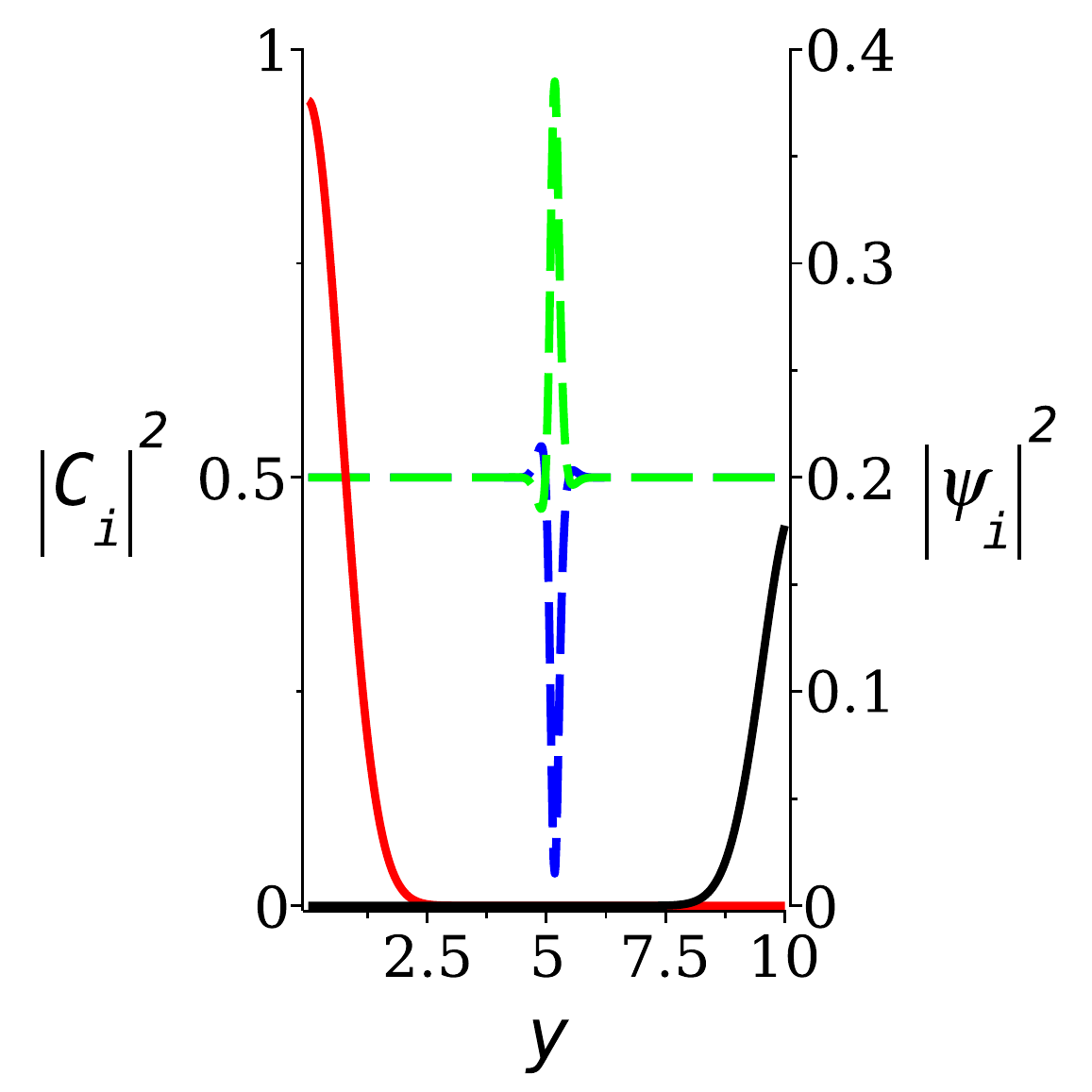}
   \end{tabular}
    \caption{XF coefficient amplitudes  $|C_i|^2$ for $i=\{L,R\}$  as green and blue dashed lines. The Born-Huang ground and excited state wavepackets are shown as red and black solid lines respectively. Snapshots of the nuclear density at different separations and the corresponding electronic coefficients in the atomic orbital basis, Eq. (\ref{eq:LocalizedPhi}). We see the presence of an interference pattern near the median point for FWPs. Initial conditions given in Table \ref{tab:ModelParamsXFPD}.}
    \label{fig:AOs}
\end{figure}

\section{Conclusions} \label{sec:summary} 
{In this paper we proposed a simple analytical model to study the numerical challenges associated with XF  of molecular wavefunctions.  The model describes the dynamics on the ground and  excited  electronic states as two diverging, non-interacting nuclear Gaussians.  The Gaussian shapes are changing much slower than their separation, which allows for the simplified treatment of moving (i.e. centers and phases depend on time) frozen (i.e. the width parameters are constant) Gaussians. }

Within the XF representation, the electronic expansion coefficients, which describe the contributions of different electronic configuration to the overall wavefunction at each nuclear coordinate, exhibit strong non-local effects, i.e. large spatial derivatives.  Their gradients approach delta functions  as $t\rightarrow\infty$, and this singular behavior  is present in both --   the localized (atomic) and molecular --  representations of the electronic coefficients.  More surprisingly, the singularity remains in the moving Lagrangian frame of the quantum trajectories, although it has a smaller prefactor compared to the stationary Eulerian frame of representing the wavefunction. {  Based on this analysis (and also  on the examination of QT dynamics for  one- and two-state cases discussed in Appendix \ref{sec:QTappendix}), one  solution is to switch to the state-specific trajectory descriptions once the nuclear wavepackets diverge when, e.g.  they reach distinct reaction channels on the Born-Huang electronic states.     We expect, that the analytic form of the singular terms identified in the model described in this works   
will  facilitate  the development of computational methodology to tackle the numerical challenges, and will help establish the limits of applicability of the XF method to the problems of non-adiabatic chemical dynamics.   Finally,  we also note, that  the model discussed in this paper  is relevant only to a discrete set of electronic states, thus the  conclusions above are limited to this case. The XF dynamics in the regime of  electronic wavepackets may be better behaved, and we intend to study this issue in the future.}




\section*{Acknowledgement}
This material is based upon work supported by  the National Science Foundation of U.S.A. under Grant No.   CHE-2308922. 
\bibliographystyle{tfq}
\bibliography{JS}

\appendix
\renewcommand{\theequation}{A.\arabic{equation}}
\renewcommand{\thefigure}{A.\arabic{figure}}

\setcounter{equation}{0}
\setcounter{figure}{0}

\section{QT dynamics on one and two electronic states}\label{sec:QTappendix}
Here we demonstrate  the quantum trajectory (QT)  dynamics  associated with one and two electronic states for the model  of Section \ref{sec:model},  somewhat  modified for illustrative purposes.  
We  present   the trajectories  assuming that two nuclear Gaussian wavepackets  evolve on distinct uncoupled electronic states,  from which an XF set of trajectories is constructed according to Eq. (\ref{eq:rhoc}). 
For comparison, we also present the  trajectories defined by the sum of the two  Gaussian wavepackets, as if they were evolving  on a single electronic state.    In the former case, 
the  XF nuclear probability density  is  given by 
\be |\psi^{XF}|^2=\bra \Psi|\Psi\ket_x=|\psi_1(y,t)|^2+|\psi_2(y,t)|^2 \label{eq:XF2state},\ee  while  in the 
latter case its counterpart is the probability density of a single nuclear wavefunction which generates the usual Bohmian trajectories,   
\be  |\psi^{Bohm}|^2=|\psi_1(y,t)+\psi_2(y,t)|^2. \label{eq:XF1state}\ee Thus, the difference between the ensuing QT ensembles  will come from the interference  of the two wavepackets in case of the single-state dynamics.  The results are presented in Fig. \ref{fig:App1}.    

First,  let us consider  the dynamics on two parabolic surfaces, 
\be V_{11}=y^2/2,~V_{22}=V_{11}+1. \label{eq:app_V1}\ee 
The constant in $V_{22}$ does not affect the QT dynamics in the two-state system, and generates a relative phase between the two wavepackets in the one-state system. 
First, we will consider two  wavepackets, $\psi_{1(2)}$,  taken as  coherent Gaussians given by  Eq. (\ref{eq:psi1}).  We will examine a symmetric and asymmetric initial conditions.  In the {\it symmetric} case,  the wavepackets are initially displaced from the minimum of $V_i$,   $q_1(0)=-1,~q_2(0)=1$,   have zero initial phase, $p_1=p_2=0$,   and their population coefficients, 
are equal, $\lambda_1=\lambda_2=1/\sqrt{2}$.  The resulting QTs are shown in Fig. \ref{fig:App1}(a,b). 
In the {\it asymmetric} case, $\psi_1$ is an eigenstate ($q_1=0,~p_1=0$, and $\psi_2$ is displaced,  $q_2(0)=1,~p_2(0)=0$;  the population coefficients  
are equal to  $\lambda_1=\sqrt{3}/{2}$ and $\lambda_2=1/2$.   The resulting QTs are shown in Fig. (\ref{fig:App1}(c,d)).   In all panels dashed red lines indicate $q_1(t)$ and $q_2(t)$;   
the blue dots and grey dashes outline the   QTs corresponding to the dynamics of  $\psi_1$ and $\psi_2$ as separate wavefunctions, while the black lines represent  the QT ensemble underlying the evolution of XF  nuclear wavefunction assuming two-electronic states (Fig. \ref{fig:App1}(a,c)), or on a single electronic state (Fig. \ref{fig:App1}(b,d)). 

In all cases the QTs describing the individual wavepacket components are smooth trajectories that are parallel to the respective wavepacket center $q_i$. (In the asymmetric case  the trajectories corresponding to $\psi_1$ are stationary.)  In the two-state scenario, the QT trajectories corresponding to the XF wavefunction, 
 are also smooth and cover nearly  (exactly for the symmetric system) the same  coordinate space as the QT ensembles of the individual wavepackets.      
In the one-state scenario, the trajectories corresponding to the full wavefunction,  $\psi=\sum_i\psi_i$, 
 travel around the nodal points at $t=\pi$  and near-nodal points at $t=\pi/2,3\pi/2$.  This singular behavior 
 arises due to the interference term, $\bra \psi_1|\psi_2\ket$, affecting the QTs in the one-state system.  
 This suggests that for a one-state dynamics,  'regularized' or approximate QT ensemble could be constructed from the trajectory weight and positions only (ignoring the phase information). Such an ensemble can be used to define  an efficient time-dependent basis, tailored to the evolving wavefunction,  for  solving the  time-dependent Schr\"odinger equation in many dimensions.

 \begin{figure}
     \centering
     \begin{tabular}{lr}\hline
 \hspace{1cm}\texttt{\large two-state dynamics}& \texttt{\large one-state dynamics~~~}\\
 \hline 
 \multicolumn{2}{c}{ {\large  symmetric $\psi(y,0)$} }\\
{\large (a)} &  {\large  (b)} \\ 
  \includegraphics[width=0.4\textwidth]{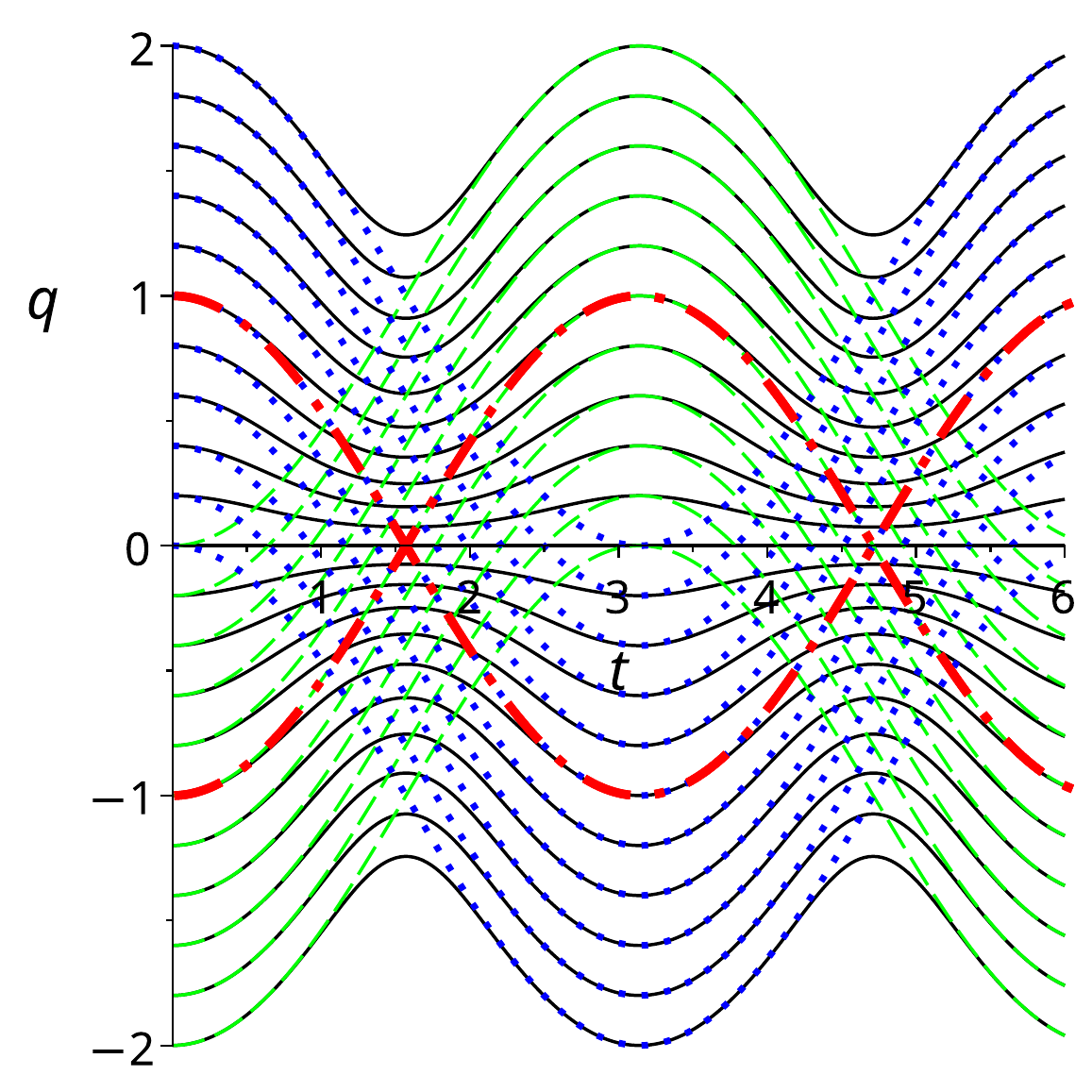} 
 &
 \includegraphics[width=0.4\textwidth]{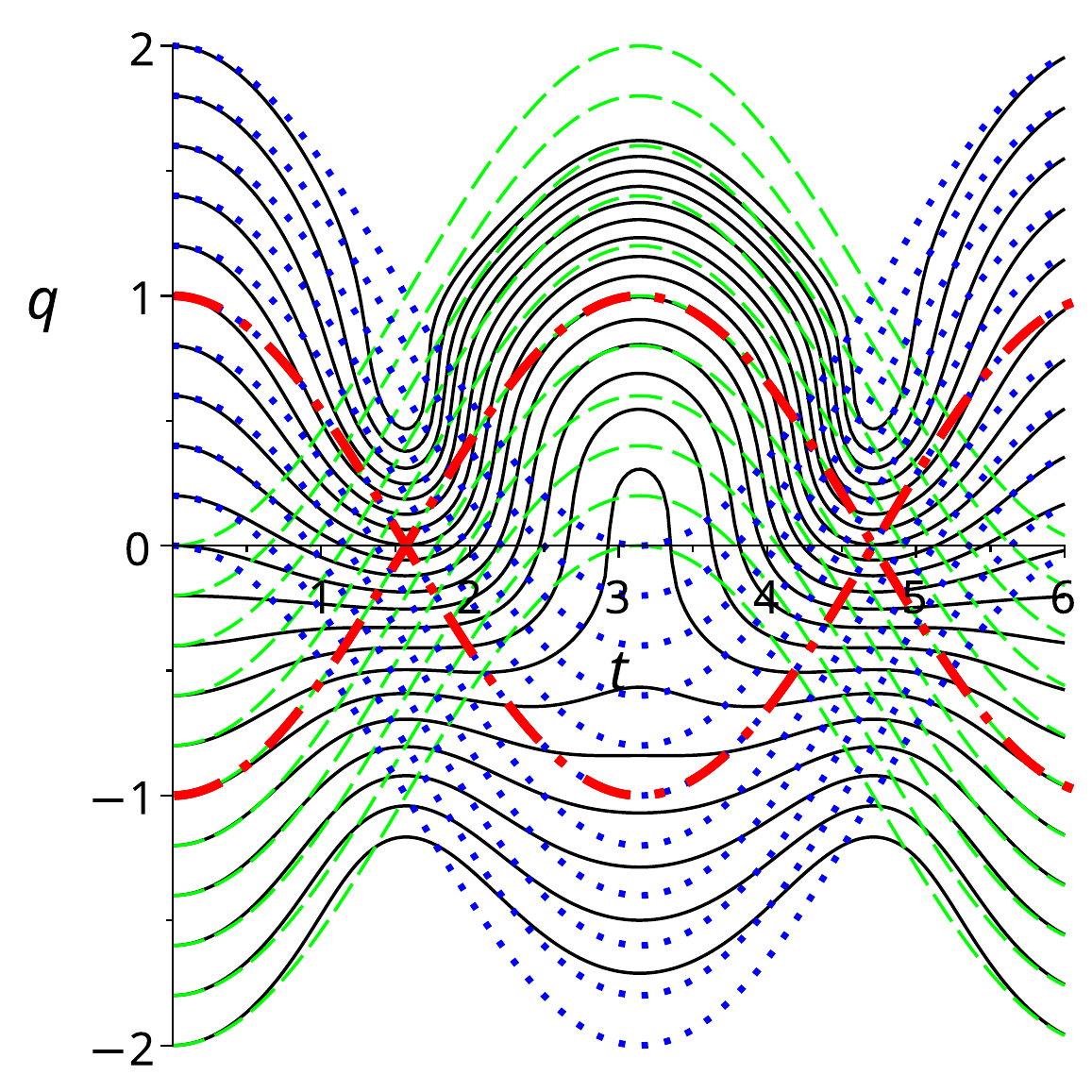}  
      \\  {\large (c)} & {\large (d)}  
   \\   \multicolumn{2}{c}{ {\large  asymmetric $\psi(y,0)$} }\\
  \includegraphics[width=0.4\textwidth]{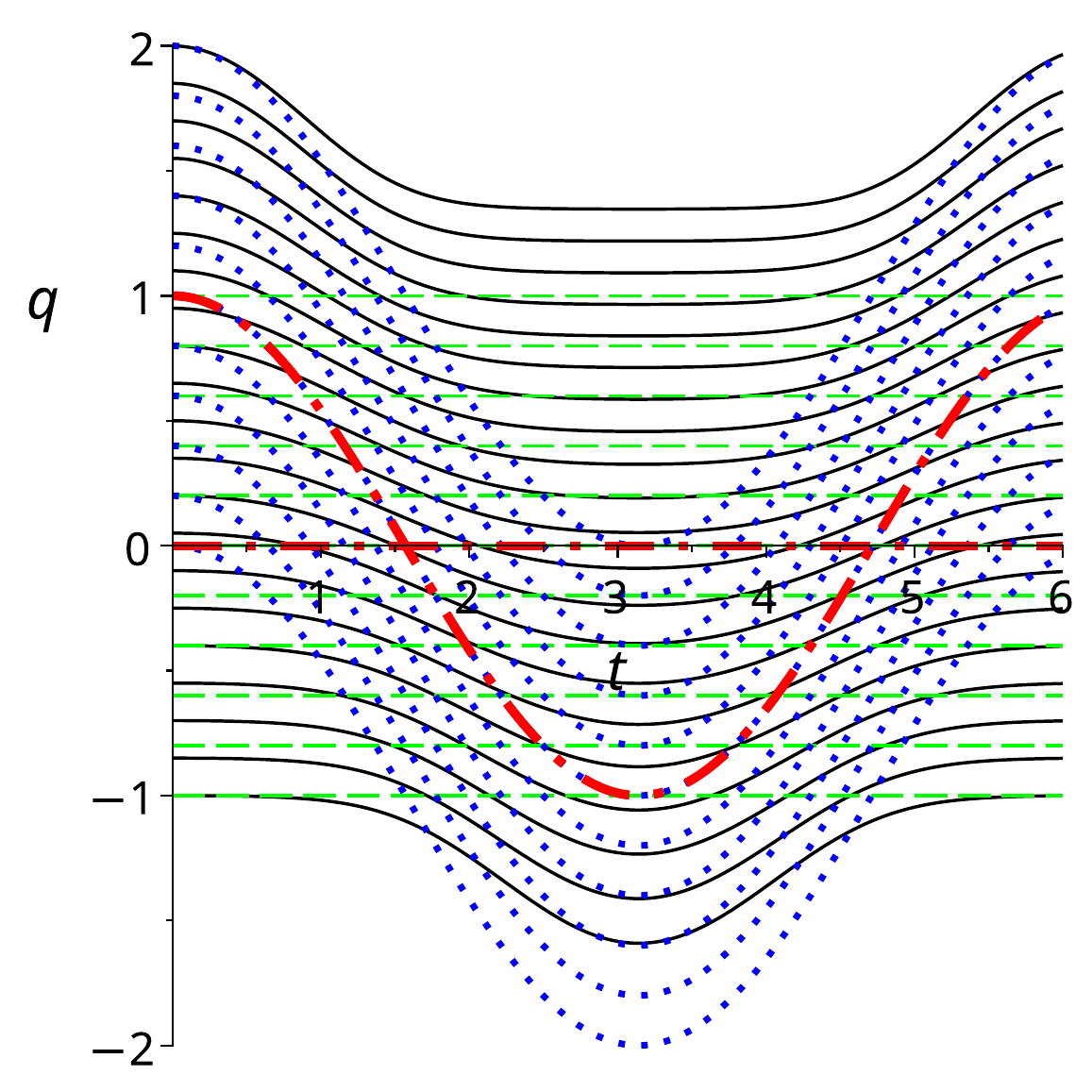}   
     & 
  \includegraphics[width=0.4\textwidth]{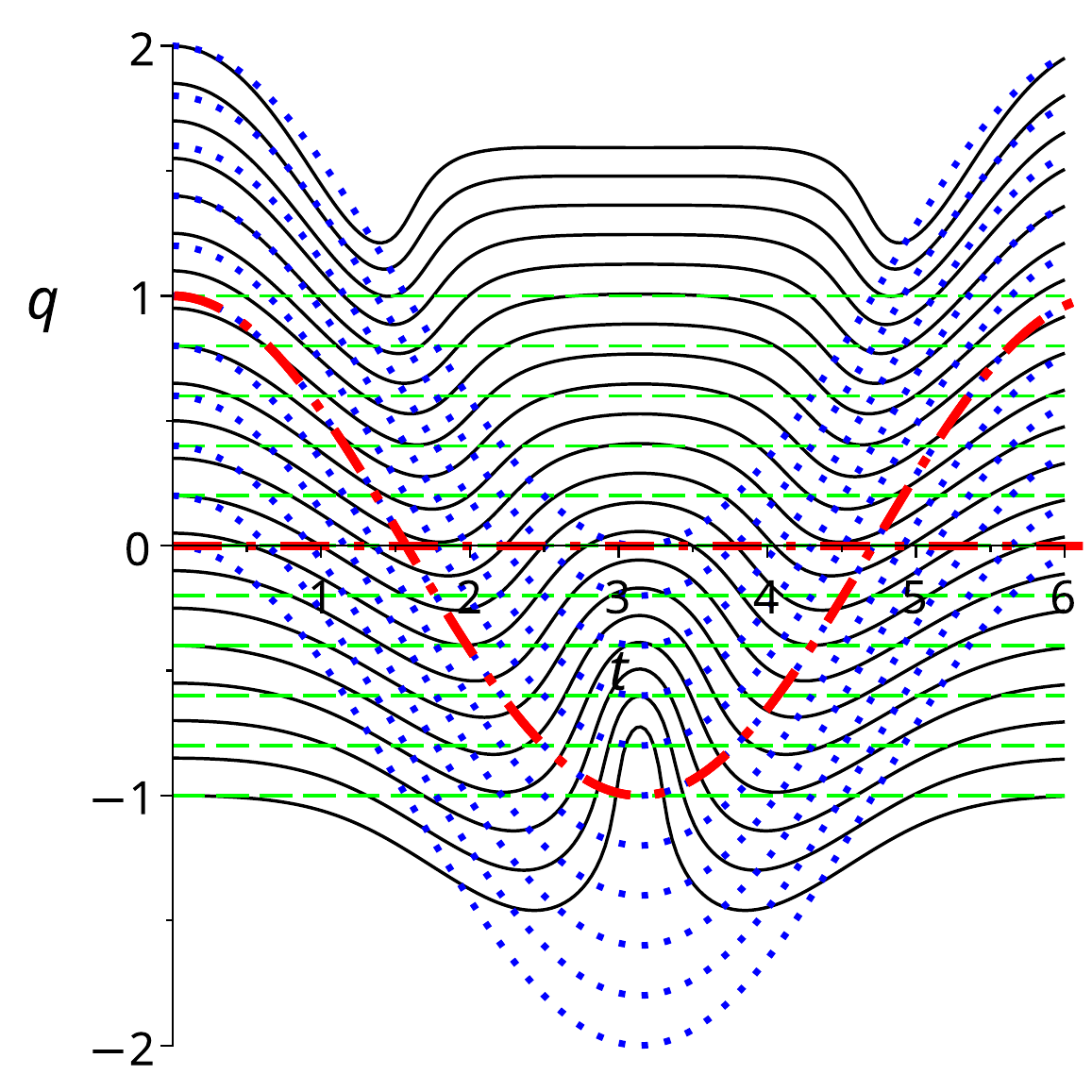}  
      \end{tabular}
\caption{Dynamics of two coherent wavepackets.
Black lines represent positions of the  QT trajectories corresponding to the dynamics of $\psi$ on two (panels (a) and (c))  and one (panels (b) and (d))  electronic states, respectively.  The  QT ensembles  for $\psi_1(\psi_2)$ are shown as  green dashed/blue dotted  lines.   The evolution of the GWP centers$q_{1(2)}$ is indicated as red dot-dashes. } 
\label{fig:App1}
     \end{figure}

 Next, let us consider dynamics of the type discussed in Section \ref{sec:TheoryAnalysis} with $V_{22}$ replaced by a linear ramp,  $V_{22}=-ky$, $k=2$ a.u.  The initial nuclear wavepackets are identical: $a_i=1$,  $q_i=-1$, $p_i=0$ a.u. The ground state wavepacket, $\psi_1$,  is a coherent Gaussian whose center oscillates as in the previous example, while the center of the second wavepacket, $\psi_2$,  moves downhill and spreads in space according to Eqs (\ref{eq:traj}-\ref{eq:gamma}). The  quantum, or Bohmian, trajectories,  associated with the motion of he individual wavepackets are shown 
 in Fig. \ref{fig:App2}(a) as families of green dashes and blue dotted lines, respectively. Red dot-dashes highlight the motion of the Gaussian centers, $q_i(t)$.  The QTs describing the XF wavefunction of Eq. (\ref{eq:XF2state}), shown as black lines, represent a 'union' of the two sets, except that for the same uniform sampling of the trajectories (used to generate all trajectory ensembles)  there are no XF trajectories  in the region of the divergence of $\psi_1$ and $\psi_2$. Fig. \ref{fig:App2}(b) illustrates the divergent region on a smaller scale;  the corresponding gradients of  the XF expansion amplitude, $\grad_y|C_1(y,t)|$ approaches the $\delta-$function as seen in Fig. \ref{fig:App2}(c).  This behavior  is  the source of the numerical challenges in implementing XF in the regime of divergent nuclear wavefunction components.  Thus, we argue that  for practical  reasons,   in this regime   the  dynamics of the nuclear components  should be 'uncoupled'.  We note however, that our  analysis is limited to a discrete  spectrum of the electronic states, and does not extend to the dynamics involving electronic wavepackets composed of large number of the electronic states. The type of singularity discussed above may be mitigated or avoided altogether in the limit of the  electronic wavepackets.       

\begin{figure}
    \centering
\begin{tabular}{lll}
    (a)  & (b) &(c)  
    \\
  \includegraphics[width=0.32\textwidth]{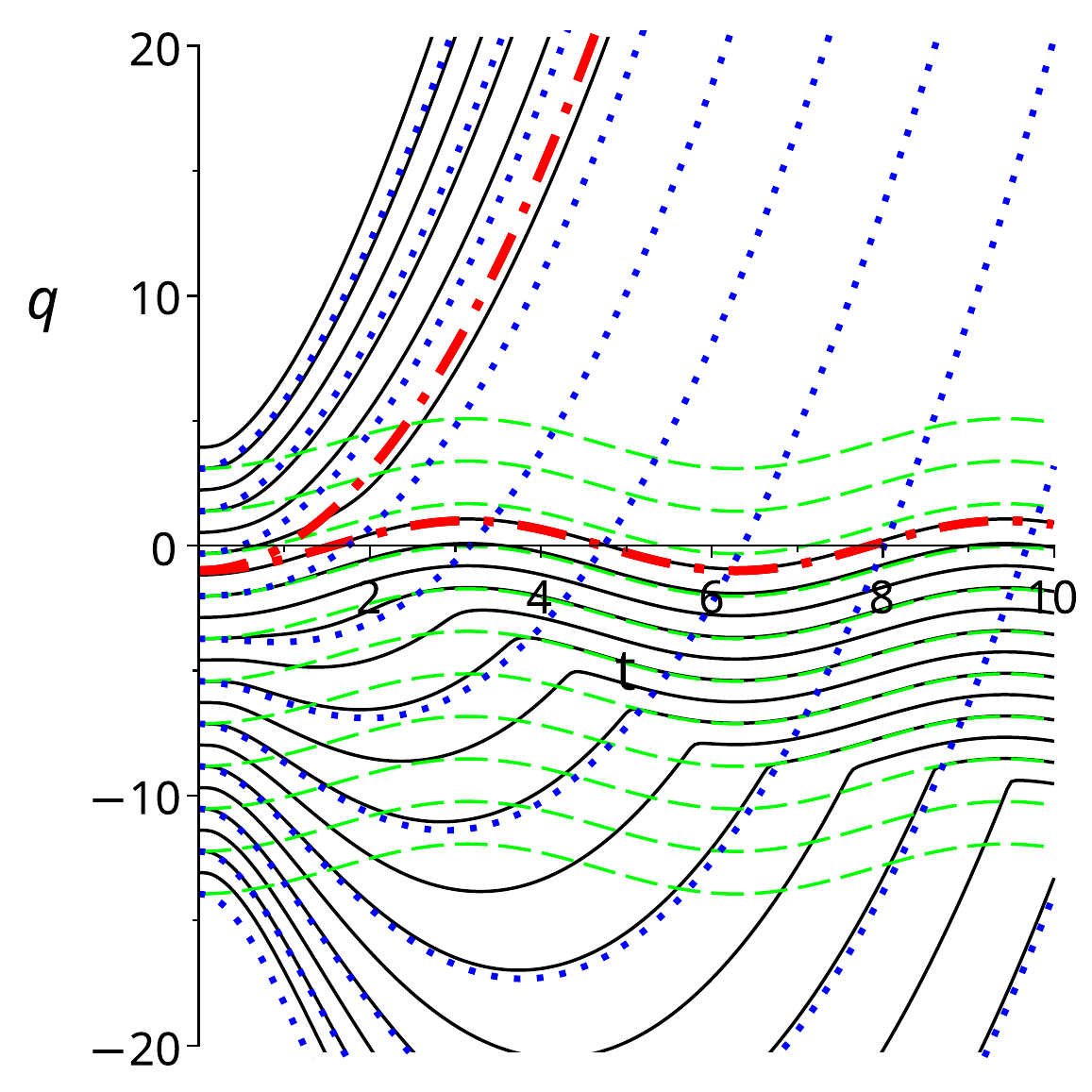} & \includegraphics[width=0.32\textwidth]{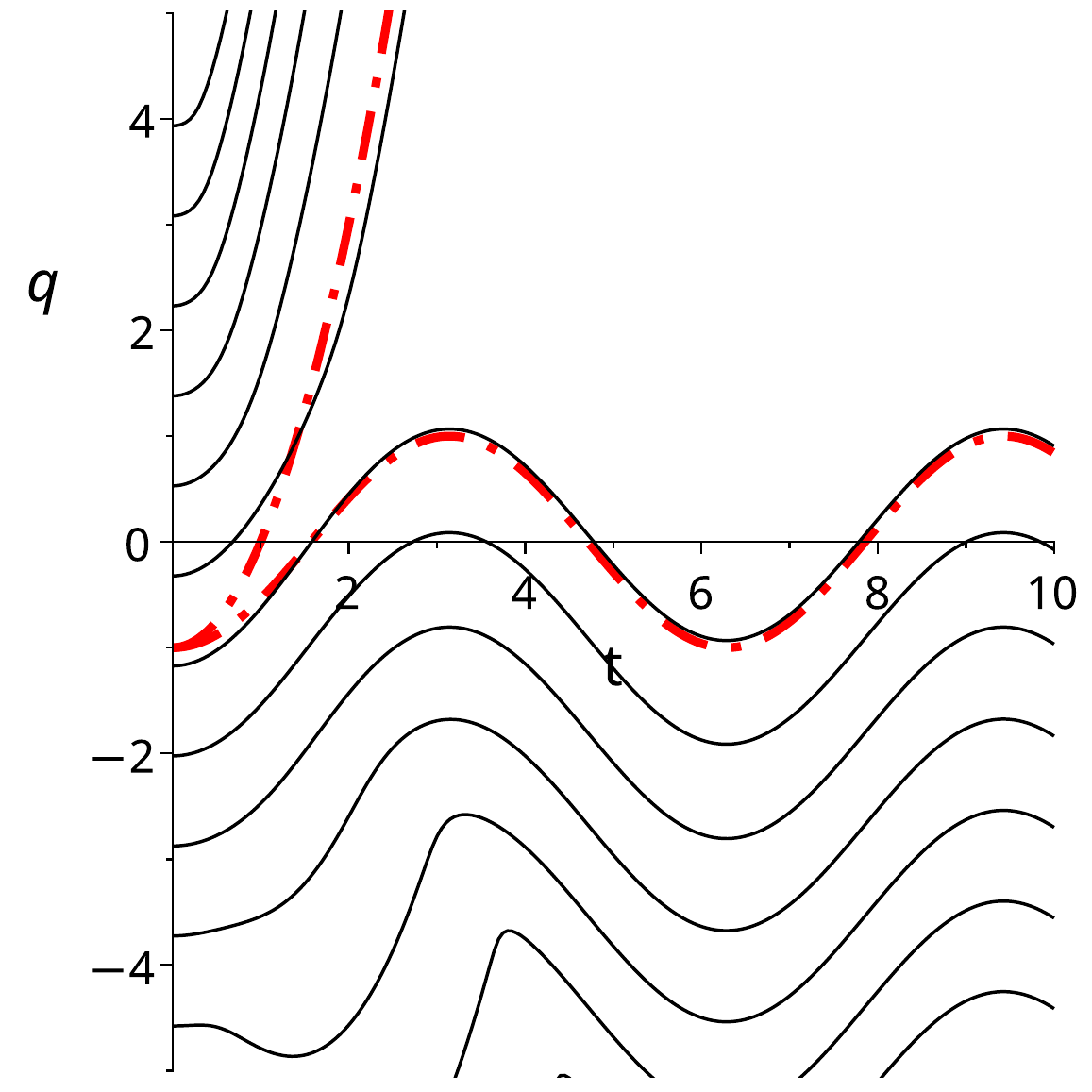} &  \includegraphics[width=0.32\textwidth]{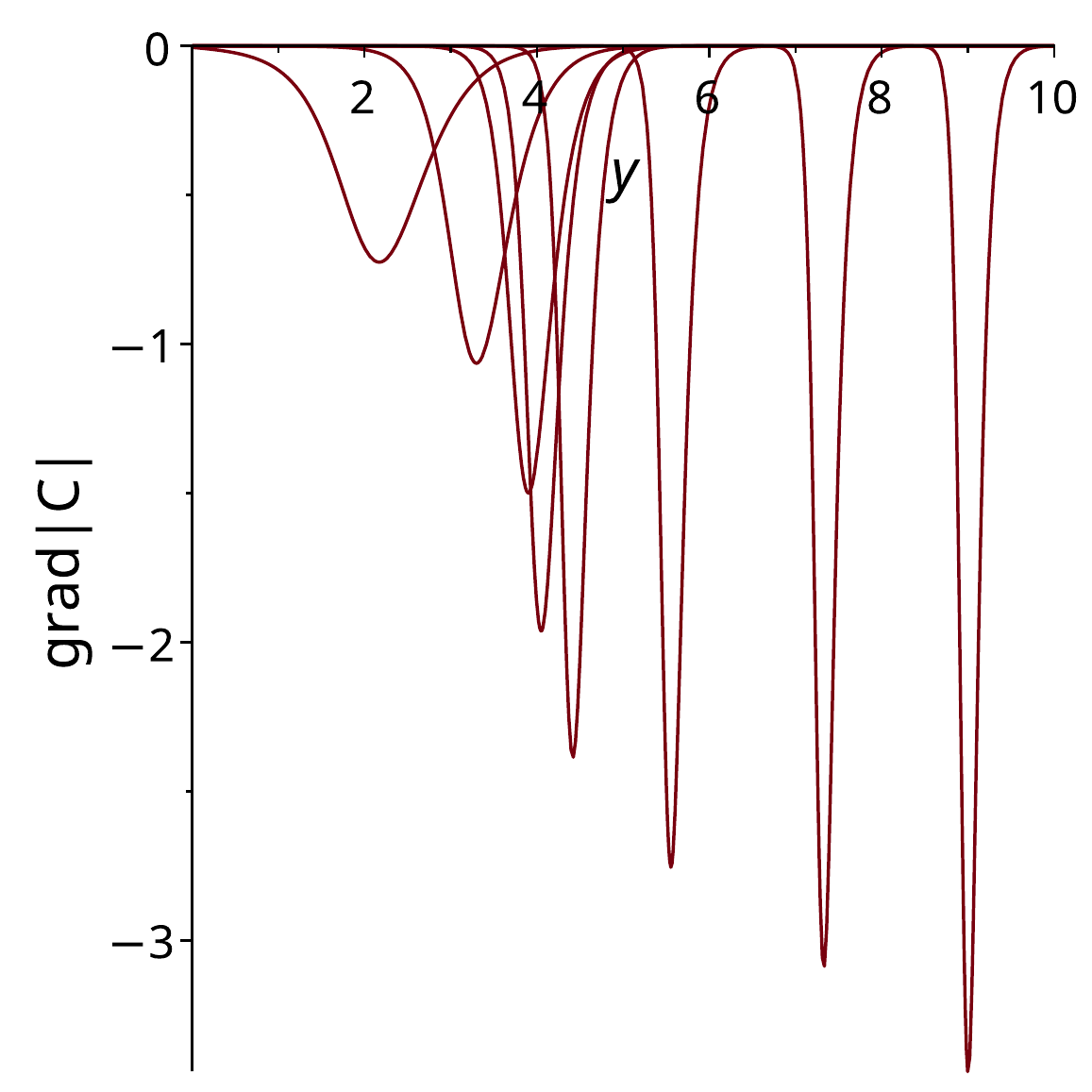}
    \end{tabular}
     \caption{Two-state dynamics  in the parabolic/linear ramp potentials.
     The parameters and initial conditions are  $\lambda_1=\sqrt{3}/2,~\lambda_2=1/2,~a=1,~q_1(0)=q_2(0)=-1, p_1(0)=p_2(0)=0$.
     (a)   The QTs of $\psi_1$ and $\psi_2$ are represented as green dashes and blue dots, respectively. The QTs of the XF wavefunction, $\psi^{XF}$,  are displayed  as black lines;  red dot-dashes represent the centers of $q_{1(2)}$ as functions of time, $t$. The  divergent region of the  XF and wavepacket center trajectories is magnified  in  panel (b). (c) The snapshots of the gradient of the XF amplitude,  $\grad_y|C_1(y,t)|$  for $t=\{2,3,4,5,6,7,8,9\}$ a.u.; the peak amplitudes increase with time, as $\grad_y|C_1(y,t)|$ approaches the $\delta-$function.}
    \label{fig:App2}
\end{figure}

\end{document}